\documentclass[nofootinbib,10pt,twocolumn]{revtex4-1}
\usepackage{amsmath,amssymb,pgf,pgfarrows,pgfnodes,float,appendix, hyperref,slashed,breakurl,graphicx}
\definecolor{Color}{rgb}{0.28, 0.24, 0.55}
\definecolor{Orange}{rgb}{1,0.38,0.11}
\hypersetup{
    colorlinks = true,
    citecolor = Orange,
    linkcolor = Color,
    urlcolor  = Orange,
}
\usepackage{makecell}
\definecolor{internationalorange}{rgb}{1.0, 0.31, 0.0}

\usepackage{graphicx}
\usepackage{subfigure}
\usepackage[margin=0.9in]{geometry}
\usepackage[compat=1.1.0]{tikz-feynman}
\usepackage{tikz}
\usepackage{amsmath}
\usepackage{tikz}
\usetikzlibrary{"arrows", "automata", "backgrounds", "calendar", "chains", "matrix", "mindmap", "patterns", "petri", "shadows", "shapes.geometric", "shapes.misc", "spy", "trees"}
\usetikzlibrary{arrows,shapes}
\usetikzlibrary{trees}
\usetikzlibrary{matrix,arrows} 				% For commutative diagram
\usepackage{bm}
\usetikzlibrary{positioning}				% For "above of=" commands
\usetikzlibrary{calc,through}				% For coordinates
\usetikzlibrary{decorations.pathreplacing}  % For curly braces
\usepackage{pgffor}							% For repeating patterns

\usetikzlibrary{decorations.pathmorphing}	% For Feynman Diagrams
\usetikzlibrary{decorations.markings}
\tikzset{
    vector/.style={decorate, decoration={snake}, draw},
	provector/.style={decorate, decoration={snake,amplitude=2.5pt}, draw},
	antivector/.style={decorate, decoration={snake,amplitude=-2.5pt}, draw},
    fermion/.style={draw=black, postaction={decorate},
        decoration={markings,mark=at position .55 with {\arrow[draw=black]{>}}}},
    fermionr/.style={draw=black, postaction={decorate},
    decoration={markings,mark=at position .55 with {\arrow[draw=black]{<}}}},
    fermioncyan/.style={draw=black, postaction={decorate},
        decoration={markings,mark=at position .55 with {\arrow[draw=cyan]{<}}}},
    fermiondif/.style={draw=black, postaction={decorate},
        decoration={markings,mark=at position .7 with {\arrow[draw=black]{>}}}},
            fermiondif2/.style={draw=black, postaction={decorate},
        decoration={markings,mark=at position .7 with {\arrow[draw=black]{<}}}},
    fermionend/.style={draw=black, postaction={decorate},
        decoration={markings,mark=at position 1 with {\arrow[draw=black]{>}}}},
    fermionuchannel2/.style={draw=black, postaction={decorate},
        decoration={markings,mark=at position .4 with {\arrow[draw=black]{>}}}},
    scalardif/.style={dashed,draw=black, postaction={decorate},
        decoration={markings,mark=at position .7 with {\arrow[draw=black]{>}}}},
    scalarend/.style={dashed,draw=black, postaction={decorate},
        decoration={markings,mark=at position 1 with {\arrow[draw=black]{>}}}},
    fermionbar/.style={draw=black, postaction={decorate},
        decoration={markings,mark=at position .55 with {\arrow[draw=black]{<}}}},
    fermionnoarrow/.style={draw=black},
    gluon/.style={decorate, draw=black,
        decoration={coil,amplitude=4pt, segment length=5pt}},
    scalar/.style={dashed,draw=black, postaction={decorate},
        decoration={markings,mark=at position .55 with {\arrow[draw=black]{>}}}},
    scalarcyan/.style={dashed,draw=black, postaction={decorate},
        decoration={markings,mark=at position .55 with {\arrow[draw=cyan]{>}}}},
    scalaruchannel1/.style={dashed,draw=black, postaction={decorate},
        decoration={markings,mark=at position .7 with {\arrow[draw=black]{>}}}},
                  scalaruchannel2/.style={dashed,draw=black, postaction={decorate},
        decoration={markings,mark=at position .4 with {\arrow[draw=black]{>}}}},
    scalarbar/.style={dashed,draw=black, postaction={decorate},
        decoration={markings,mark=at position .55 with {\arrow[draw=black]{<}}}},
    scalarnoarrow/.style={dashed,draw=black},
    electron/.style={draw=black, postaction={decorate},
        decoration={markings,mark=at position .55 with {\arrow[draw=black]{>}}}},
	bigvector/.style={decorate, decoration={snake,amplitude=4pt}, draw},
}

\NewDocumentCommand\semiloop{O{black}mmmO{}O{above}}
{%
\draw[#1] let \p1 = ($(#3)-(#2)$) in (#3) arc (#4:({#4+180}):({0.5*veclen(\x1,\y1)})node[midway, #6] {#5};)
}

\tikzstyle{block} = [draw, rectangle, 
    minimum height=3em, minimum width=6em]

\usepackage{enumitem}
\usepackage{tikz}
\usetikzlibrary{fit}
\tikzset{%
  highlight/.style={rectangle,rounded corners,color=granate,draw,text opacity =1,
    fill opacity=0.5,thick,inner sep=0pt}
}

\usepackage{xparse}
\NewDocumentCommand\loopv{O{black}mmmO{}O{above}}
{%
\draw[#1] let \p1 = ($(#3)-(#2)$) in (#3) arc (#4:({#4+360}):({0.5*veclen(\x1,\y1)})node[midway, #6] {#5};)
}

\usepackage{placeins}

\tikzset{
    cross/.pic = {
    \draw[rotate = 45] (-#1,0) -- (#1,0);
    \draw[rotate = 45] (0,-#1) -- (0, #1);
    }
}

\tikzset{
    square/.style={%
        draw=none,
        circle,
        append after command={%
            \pgfextra \draw[#1] (\tikzlastnode.north-|\tikzlastnode.west) rectangle 
                (\tikzlastnode.south-|\tikzlastnode.east);\endpgfextra}
    },
    square/.default=black
}

\tikzstyle{block} = [draw, rectangle, 
    minimum height=3em, minimum width=6em]

\usepackage{amsmath}
\usepackage{todonotes}
\usepackage[utf8]{inputenc}
\usepackage[T1]{fontenc} 
\usepackage{array}
\usepackage{bbm}
\usepackage{soul}	
\usepackage{xcolor}
\usepackage{slashed}
\usepackage{epsfig}
\usepackage{color}
\usepackage{slashed}
\usepackage{comment}
\usepackage{epstopdf}
\usepackage{xspace}
\usepackage{mathrsfs}
\usepackage[euler]{textgreek}
\usepackage{amsmath}
\usepackage{amsthm}
\usepackage{amsfonts}
\usepackage{amssymb}
\usepackage{graphicx}
\usepackage[font={small}]{caption}
\usepackage{subcaption}
\usepackage{float}
\usepackage{multirow}
\usepackage[hang, flushmargin,bottom]{footmisc} 
\usepackage{placeins}
\usepackage{enumitem}
\usepackage{slashed}
\usepackage{bbm}
\usepackage{appendix}
\usepackage{tabularx}
\usepackage{siunitx}

\usepackage{titlesec}
\titleformat{\chapter}[display]
  {\normalfont\LARGE\bfseries}
  {\chaptertitlename\ \thechapter}{5pt}{\LARGE}
  \titlespacing*{\chapter}{0pt}{-20pt}{35pt}
\usepackage{bigstrut}
\usepackage{pst-node}
\usepackage{pstricks}
\usepackage{physics}
\usepackage{mathtools}
\usepackage{setspace}
\usepackage{fancyhdr}
\usepackage[makeroom]{cancel}
\usepackage{feyn}
\newcommand{\be}{\begin{equation}}
\newcommand{\ee}{\end{equation}}
\newcommand{\bes}{\begin{equation*}}
\newcommand{\ees}{\end{equation*}}

\newcommand{\e}{\text{e}}

\usepackage{hyperref}% http://ctan.org/pkg/hyperref
\hypersetup{%
  colorlinks = true,
  linkcolor  = blue,
  citecolor  = blue,
  urlcolor   = blue,
}
\usepackage{soul}
\usepackage{xpatch}
\makeatletter
\xpretocmd{\todo}{\@bsphack}{}{}
\xapptocmd{\todo}{\@esphack}{}{}
\makeatother

\newcommand{\beq}{\begin{equation}}
\newcommand{\eeq}{\end{equation}}

\usepackage{tikz}

\definecolor{green}{HTML}{008000}
\definecolor{goldenrod}{HTML}{DAA520}
\definecolor{magenta}{HTML}{FF00FF}
\definecolor{silver}{HTML}{C0C0C0}
\definecolor{indigo}{HTML}{4B0082}
\definecolor{skyblue}{HTML}{87CEEB}
\definecolor{darkgoldenrod}{HTML}{B8860B}
\definecolor{orange}{HTML}{FFA500}
\definecolor{yellow}{HTML}{FFFF00}
\definecolor{saddlebrown}{HTML}{8B4513}
\definecolor{blue}{HTML}{0000FF}
\definecolor{turquoise}{HTML}{40E0D0}
\definecolor{yellow}{HTML}{FFFF00}
\definecolor{white}{HTML}{FFFFFF}
\definecolor{whitesmoke}{HTML}{F5F5F5}
\definecolor{hotpink}{HTML}{FF69B4}
\newcommand{\myComment}[1]{}

\begin{document}
\title{\Large{Lepton Flavor Violation and Local Lepton Number}}
\author{Hridoy Debnath, Pavel Fileviez P{\'e}rez}
\affiliation{
Physics Department and Center for Education and Research in Cosmology and Astrophysics (CERCA), Case Western Reserve University, Cleveland, OH 44106, USA }
\email{hxd253@case.edu, pxf112@case.edu}
\date{\today}

\begin{abstract}
We investigate the predictions for lepton number violating processes within the minimal theory of neutrino masses based on the spontaneous breaking of local lepton number. In this framework, the symmetry is broken at the low scale, leading to the existence of a viable dark matter candidate. The new fermions required for anomaly cancellation mediate lepton number violating processes at the one-loop level. We present a detailed calculation of the most relevant processes, including $\mu \to e \gamma$, $\mu \to 3 e$, and $\mu \to e$ conversion in nuclei. The regions of parameter space excluded by current experimental bounds are identified, and we emphasize the interplay between collider observables and charged lepton flavor violating signatures as a key test of this minimal theory of neutrino masses.
\end{abstract}

\maketitle

%%%%%%%%%%%%%%%%%%%%%%%%%
\section{INTRODUCTION}
%%%%%%%%%%%%%%%%%%%%%%%%%
The discovery of neutrino oscillations has established that lepton flavor is not conserved in nature, providing the first confirmed example of physics beyond the Standard Model (SM) of particle physics. While neutrino flavor violation is now firmly established, charged lepton flavor violation (LFV) remains unobserved. Within the SM with massive Dirac neutrino masses, the rates for processes such as $\mu \to e \gamma$, $\mu \to 3 e$, or $\mu \to e$ conversion in nuclei are highly suppressed, far below any foreseeable experimental sensitivity. Clearly, any observation of LFV in the charged lepton sector would be a clear signal of new physics. The current experimental bounds on LFV processes are very strong, one has 
for example that ${\rm{BR}}(\mu \to e \gamma) < 1.5 \times 10^{-13}$~\cite{afanaciev2025newlimitmuegammadecay}, with the possibility to improve it in almost an order of magnitude. The Mu2e experiment at Fermilab will play a very important role due to the possibility to achieve a very strong bound on $\mu \to e$ conversion, ${\rm{BR}}(\mu \ \rm{Al} \to e \ \rm{Al}) \sim 10^{-17}$~\cite{COMET:2025sdw}. For reviews on LFV processes, see, for example Refs.~\cite{Bernstein:2013hba,Calibbi:2017uvl,Davidson:2022jai,Davidson:2022nnl}.

In theories for physics beyond the SM one can have large predictions for LFV processes. In particular, in theories of neutrino masses one can expect strong limits from LFV experimental bounds. 
We know that neutrinos are massive particles but their origin and nature is unknown. 
As is well-known, the key symmetry one needs to understand to distinguish between Majorana and Dirac neutrinos is the total lepton number. The total lepton number is defined as $\ell=\ell_e+\ell_\mu+\ell_\tau$, with $\ell_i$ being the lepton number for a given family. This symmetry is conserved only at the classical level in the SM, broken at the quantum level by $SU(2)$ instantons~\cite{tHooft:1976rip}. 

Theories based on gauging the total lepton number offer a simple and predictive framework for addressing the origin of neutrino masses and dark matter~\cite{FileviezPerez:2010gw,FileviezPerez:2011pt,Duerr:2013dza,FileviezPerez:2014lnj,FileviezPerez:2024fzc}. In the minimal realization~\cite{FileviezPerez:2024fzc}, anomaly cancellation requires the introduction of a small number (only four) of new fermionic fields, one of which can serve as a viable dark matter candidate. In this context, the spontaneous breaking of the local lepton number symmetry generates masses for these new states and predicts a new neutral gauge boson, which mass 
is tightly connected to the scale of symmetry breaking. One of the most interesting features of these theories is that one finds an upper bound on the symmetry breaking scale using the experimental constraints on the dark matter relic density~\cite{Debnath:2025rbu}.
Since this upper bound is in the multi-TeV, $M_{Z_\ell} \lesssim 30$ TeV, one can hope to test the origin of neutrino masses at colliders.

In this article, we investigate in detail the theoretical predictions for lepton flavor violating processes within the minimal theory of local lepton number~\cite{FileviezPerez:2024fzc}. We focus on radiative decays such as $\mu \to e \gamma$ and $\tau \to \mu \gamma$, three-body decays like $\mu \to 3e$, and $\mu \to e$ conversion in nuclei. We show how current and future experimental searches strongly constrain the parameter space of the theory and establish correlations between LFV bounds, dark matter phenomenology, and collider signatures. In this way, LFV processes provide a complementary and powerful probe of the minimal local lepton number theory of neutrino masses. 
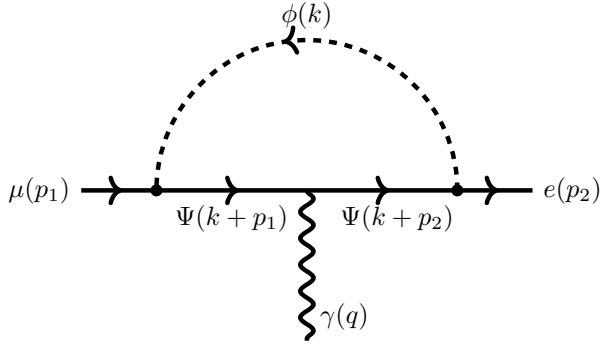
\begin{figure}[h]
\centering
\begin{tikzpicture}[line width= 1.7 pt,node distance=1 cm and 1 cm]
\coordinate[label=left:$\mu(p_1)$](a);
\coordinate[right = 1 cm of a](v1);
\coordinate[right = 4 cm of v1](v2);
\coordinate[right = 2 cm of v1](aux);
\coordinate[above = 2 cm of aux,label=$\phi(k)$](phi);
\coordinate[right = 1.0 cm of v2,label=right:$e(p_2)$](e);
\coordinate[right = 2 cm of v1](psi);
\coordinate[below = 2 cm of psi](psi2);
\coordinate[right = 1 cm of v1](psi3);
\coordinate[right = 3.2 cm of v1](psi4);
\coordinate[right = 2.5 cm of v1](g1);
\coordinate[below = 0.7 cm of psi3,label=$\Psi(k+p_1)$](psi5);
\coordinate[below = 0.7 cm of psi4,label=$\Psi(k+p_2)$](psi6);
\coordinate[below = 2 cm of g1,label=$\gamma(q)$](gamma);
\draw[fill=black](v1) circle (.05cm);
\draw[fill=black](v2) circle (.05cm);
\draw[fermion](a)--(v1);
\draw[fermion](v1)--(psi);
%\draw[fermionnoarrow](v1)--(v2);
\draw[fermion](psi)--(v2);
\draw[fermion](v2)--(e);
\draw[vector](psi) -- (psi2);
\semiloop[scalar]{v1}{v2}{0};
\end{tikzpicture}
\caption{One-loop Feynman diagram for $\mu \to e \gamma$.}
\label{mueg}
\end{figure}

This article is organized as follows: in Section~\ref{theory} we discuss the minimal theory of neutrino masses based on local lepton number, while in Section~\ref{LFV} we show the predictions for lepton flavour violating processes. We also discuss the correlation between the collider and lepton number violating signatures. In Section~\ref{summary]} we summarize our main findings. In the appendices we provide the full calculation of the main lepton number violating processes and list the corresponding experimental bounds.
%
%%%%%%%%%%%%%%%%%%%%%%%%%%%%%%%%%%%%%%%%%%%%%%%%%%
\section{LOCAL LEPTON NUMBER}
\label{theory}
%%%%%%%%%%%%%%%%%%%%%%%%%%%%%%%%%%%%%%%%%%%%%%%%%%
Simple theories in which the total lepton number is a local gauge symmetry have been proposed~\cite{FileviezPerez:2010gw,FileviezPerez:2011pt,Duerr:2013dza,FileviezPerez:2014lnj,FileviezPerez:2024fzc}. 
These theories are based on the gauge group
\begin{equation}
SU(3)_C \otimes SU(2)_L \otimes U(1)_Y \otimes U(1)_\ell.  \nonumber
\end{equation}
In the minimal theory proposed in~\cite{FileviezPerez:2024fzc} the gauge anomalies are canceled with only four fermionic representations plus the right-handed neutrinos. In this context, the new fermionic fields are:
\begin{eqnarray}
\Psi_L &\sim& ({\bf{1}},{\bf{1}},-1,3/4), \    \Psi_R \sim ({\bf{1}},{\bf{1}},-1,-3/4), \nonumber \\
\chi_L &\sim& ({\bf{1}},{\bf{1}},0,3/4), \ 
{\rm{and}} \ \rho_L \sim ({\bf{1}},{\bf{3}},0,-3/4). \nonumber
\end{eqnarray}
\begin{figure}[t]
        \includegraphics[width=0.45\textwidth]{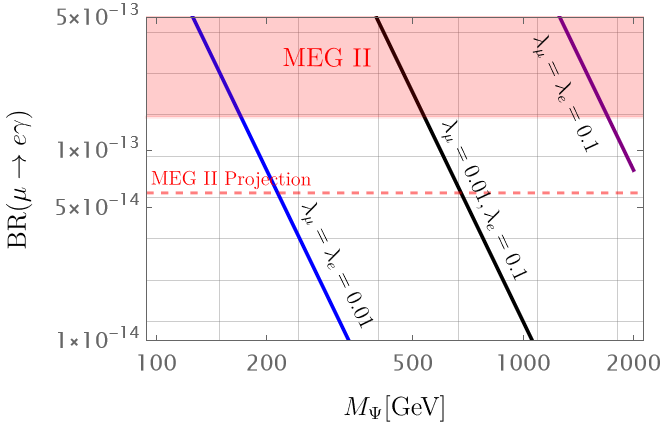}
        \includegraphics[width=0.45\textwidth]{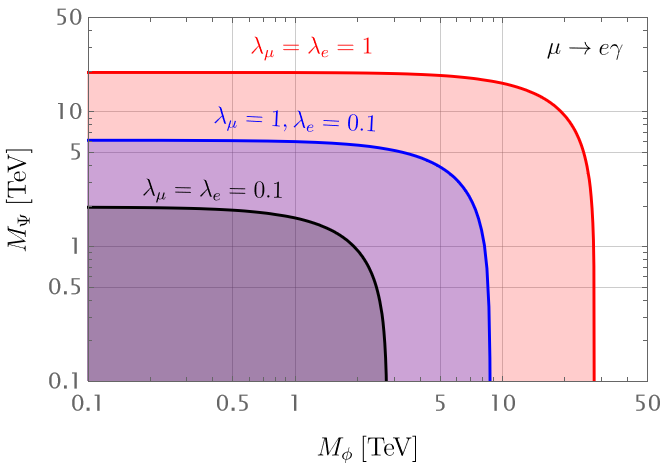}
\caption{In the upper panel we show the branching ratio for $\mu \to e \gamma $ vs. $M_{\Psi}$ for different values of $\lambda_e$ and  $\lambda_\mu$. The red region is excluded by the MEG II experiment bounds~\cite{afanaciev2025newlimitmuegammadecay} and the dashed red line shows the projected MEG II bound~\cite{Baldini_2021}. Here we assume $M_\Psi=2 M_\phi$. In the lower panel we show the allowed parameter space in the $M_\Psi -M_\phi$ plane for different values of $\lambda_e$ and $\lambda_\mu$. The shaded regions are excluded by MEG II~\cite{afanaciev2025newlimitmuegammadecay}. }
\label{fig:mutoegamma}
 \end{figure}
We can generate masses for the new fields using the new Yukawa interactions:
\begin{eqnarray}
- \mathcal{L} &\supset& \lambda_\rho {\rm{Tr}} (\rho_L^T C \rho_L) S + \lambda_\Psi \bar{\Psi}_L \Psi_R S + \lambda_\chi \chi_L^T C \chi_L S^* \nonumber \\
&+& \rm{H.c.},   
\end{eqnarray}
where $S\sim ({\bf{1}},{\bf{1}},0,3/2)$ is a new scalar field responsible for spontaneous breaking of $U(1)_\ell$.  
The SM leptons have the following Yukawa interactions:
\begin{eqnarray}
- \mathcal{L} \supset Y_\nu  \bar{\ell}_L i \sigma_2 H^* \nu_R  + 
y_e  \bar{\ell}_L H e_R
+\rm{H.c.}
\end{eqnarray}
Notice that only one extra scalar field, $S$, with lepton number $3/2$ is needed to generate masses for the new fermions, and the neutrinos are predicted to be Dirac fermions. The scalar fields can be written as 
\begin{eqnarray}
H&=&\begin{pmatrix}
h^+\\
\frac{1}{\sqrt{2}}(v_{0} + h_0) e^{i \sigma_0/v_0}
\end{pmatrix}, 
\end{eqnarray}
and
\begin{eqnarray}
S &=& \frac{1}{\sqrt{2}}\left(v_{S} + h_S \right) e^{i \sigma_S/v_S}.
\label{S-filed}  
\end{eqnarray}
After symmetry breaking, $v_0\neq0$ and $v_S\neq0$, the physical neutral CP-even Higgses are defined as
\begin{eqnarray}
h &=& h_0 \cos \theta_\ell - h_S \sin \theta_\ell,\\
h_\ell &=& h_0 \sin \theta_\ell  + h_S \cos \theta_\ell.
\end{eqnarray}
 This theory predicts a new neutral gauge boson that couples mainly to the SM leptons and the new fermions needed for anomaly cancelation. The new gauge boson acquires mass after the $U(1)_\ell$ gauge symmetry is broken, and its mass is given by
\begin{equation}
    M_{Z_\ell}=\frac{3}{2} g_\ell v_S.
\end{equation}
The $\Psi_L$ field decays at the renormalizable level if one has the extra scalar field $\phi \sim ({\bf{1}},{\bf{1}},0,-1/4)$ and the following interactions~\cite{FileviezPerez:2024fzc}:
\begin{equation}
- \mathcal{L} \supset \lambda_e \overline{\Psi}_L e_R \phi + \lambda_R \overline{\chi}_L \nu_R \phi \ + \ {\rm{H.c.}}   
\label{LFV-Yukawa}
\end{equation}
 \begin{figure}
\includegraphics[width=0.4\textwidth]{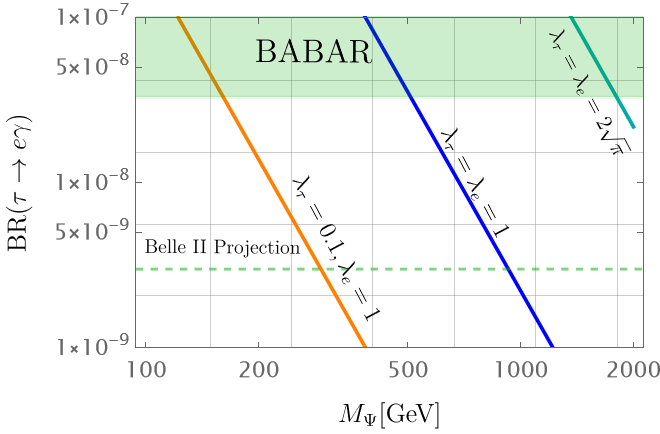}
\includegraphics[width=0.4\textwidth]{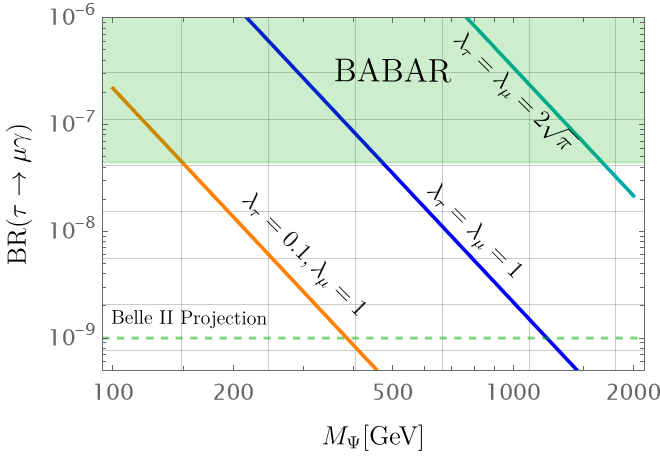}
    \caption{As in Fig.~\ref{fig:mutoegamma} we show the shaded region excluded by the BABAR experiment~\cite{Aubert_2010}. In the upper panel, we show the impact of the experimental bounds on the ${\rm{BR}}(\tau \to e \gamma)$, while in the lower panel, we show the results for ${\rm{BR}}(\tau \to \mu \gamma)$. The green dashed line shows the projected Belle II sensitivity~\cite{Belle-II:2018jsg}.}
    \label{fig:tautoegamma}
 \end{figure}
These interactions are crucial to understanding the predictions for lepton-flavor violation in this theory. 

After symmetry breaking, one predicts two extra electrically charged physical fermions: 
\begin{equation}
\Psi^-=\Psi^-_L + \Psi^-_R \ 
{\textrm{and}} \ 
\rho^-=\rho_L^-+(\rho_L^+)^C, \nonumber
\end{equation}
and two neutral Majorana fields: 
\begin{equation}
\chi=\chi_L + (\chi_L)^C \ {\textrm{and}} \ \rho^0=\rho_L^0+(\rho_L^0)^C.
\nonumber
\end{equation}
Their masses are given by 
\begin{eqnarray}
M_{\rho^0} &=& \sqrt{2} \lambda_\rho v_S, \
M_{\rho^-} = M_{\rho^0} + \delta M, \\
M_\chi &=& \sqrt{2} \lambda_\chi v_S, \ {\rm{and}} \
M_\Psi = \lambda_\Psi \frac{v_S}{\sqrt{2}}.
\end{eqnarray}
The new physical charged fermion $\rho^-$ can decay into neutral field $\rho^0$, and a charged pion $\pi^-$, due to the fact that a very small splitting is generated at the one-loop level~\cite{Cirelli:2005uq}.
Recently, we have studied some of the phenomenological aspects of this theory. Here, we provide a summary of these results:
\begin{itemize}
\item {\textit{Upper bound on the Symmetry-Breaking Scale}}: This theory predicts a cold dark matter candidate from anomaly cancellation. In Ref.~\cite{Debnath:2025rbu} we provided a detailed analysis of the dark matter candidates, and investigated the predictions for direct and indirect detection experiments. One finds that the relic density constraints give us an upper bound on the symmetry-breaking scale in the multi-TeV region, i.e. $M_{Z_\ell} \lesssim 30$ TeV~\cite{Debnath:2025rbu}. Therefore, one can be optimistic about the possibility to test this theory in the near future.
\item {\textit{Neutrino Masses}}: The minimal theory proposed in Ref.~\cite{FileviezPerez:2024fzc} predicts Dirac neutrinos, see also the study in Ref.~\cite{Debnath:2025rbu} for the case of Majorana neutrinos.
\item  {\textit{Collider Signatures}}: This theory predicts very interesting collider signatures. One can have the following distinctive signatures at the Large Hadron Collider:
\begin{itemize}
\item a) the $pp \to \rho^+ \rho^-\to \rho^0 \rho^0 \pi^+ \pi^-$ channel gives rise
  to two ``kinked'' charged tracks because the $\rho^\pm$ are long-lived with a decay length of order five centimeters.
  Here, the $\rho^0$ field is long-lived and neutral, giving rise to missing energy signatures.
\item b) charged track and missing energy, 
  one can have associated production $pp \to \rho^0 \rho^- \to \rho^0 \rho^0 \pi^-$, and since the $\rho^-$ field
is long lived, one has a charged track until this field decays into $\rho^0$ and a pion.
  The $\rho^0$ field is neutral and long-lived. 
\item c) Pair production, $ p p \to \Psi^+ \Psi^-    \to \ell_i^{+} \ell_j^{-} \phi^* \phi$, gives rise to two charged leptons and missing energy. 
  The field $\phi$ is neutral and long-lived in this context. 
\end{itemize}
  These collider signatures were studied in Ref.~\cite{Butterworth:2025asm} in the context of a different model. However, the studies in Ref.~\cite{Butterworth:2025asm} where we assume the production of the new fermions via the SM gauge bosons can be used for the model we investigate in this paper.
  In a future publication we will discuss these signatures at a future Muon-collider.
\end{itemize}
%%%%%%%%%%%%%%%%%%%%%%%%%%%%%%%%%%%%%%
\section{LEPTON FLAVOUR VIOLATION}
\label{LFV}
%%%%%%%%%%%%%%%%%%%%%%%%%%%%%%%%%%%%%%
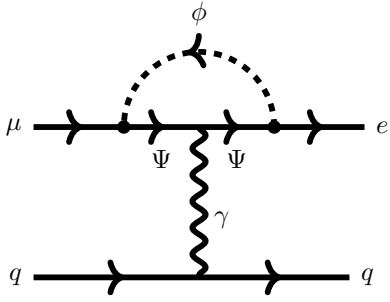
\begin{figure}[t]
\begin{eqnarray*}
\begin{gathered}
\begin{tikzpicture}[line width= 2.2 pt,node distance=1 cm and 1 cm]
\coordinate[label=left:$\mu$](a);
\coordinate[right = 1.2 cm of a](v1);
\coordinate[right = 2 cm of v1](v2);
\coordinate[right = 1 cm of v1](aux);
\coordinate[above = 1.2 cm of aux,label=$\phi$](phi);
\coordinate[right = 1.2 cm of v2,label=right:$e$](e);
\coordinate[right = 1 cm of v1](psi);
\coordinate[below = 2 cm of psi](psi2);
\coordinate[left = 2.2 cm of psi2,label=left:$q$](q2);
\coordinate[right = 2.0 cm of psi2,label=right:$q$](q3);
\coordinate[right = 0.5 cm of v1](psi3);
\coordinate[right = 1.5 cm of v1](psi4);
\coordinate[right = 1.3 cm of v1](g1);
\coordinate[below = 0.7 cm of psi3,label=$\Psi$](psi5);
\coordinate[below = 0.7 cm of psi4,label=$\Psi$](psi6);
\coordinate[below = 1.5 cm of g1,label={$\gamma$}](gamma);
\draw[fill=black](v1) circle (.05cm);
\draw[fill=black](v2) circle (.05cm);
\draw[fermion](a)--(v1);
\draw[fermion](v1)--(psi);
%\draw[fermionnoarrow](v1)--(v2);
\draw[fermion](psi)--(v2);
\draw[fermion](v2)--(e);
\draw[vector](psi) -- (psi2);
\draw[fermion](q2)--(psi2);
\draw[fermion](psi2)--(q3);
\semiloop[scalar]{v1}{v2}{0};
\end{tikzpicture}
\end{gathered} 
\quad   \! 
\end{eqnarray*}
 \caption{Feynman diagram of the photon contribution for $\mu $ to e conversion. }
    \label{mutoe1}
\end{figure} 
\begin{figure}[h]
    \includegraphics[width=0.4\textwidth]{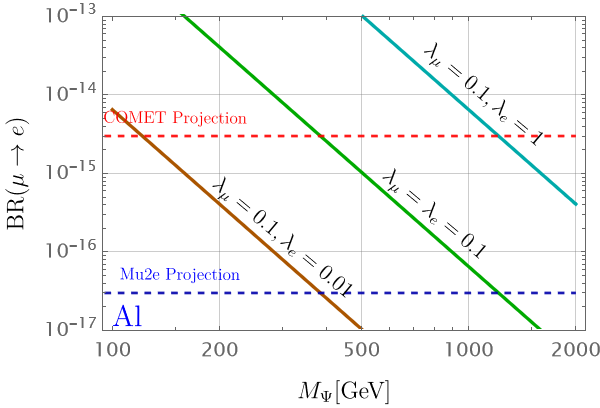}        \includegraphics[width=0.45\textwidth]{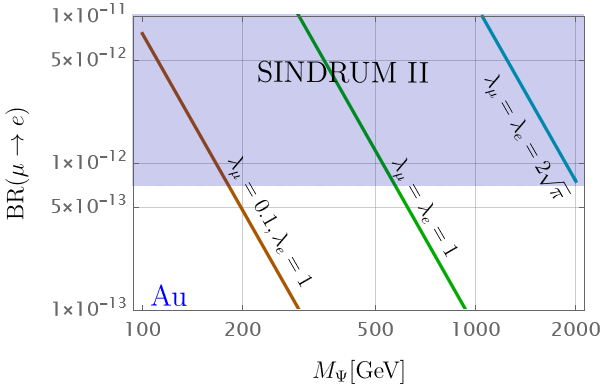}
    \caption{In the upper-panel we show the prediction for the branching ratio of $\mu \to e$ conversion in Al. The red and blue dashed lines show the projected bounds by the COMET and Mu2e experiments~\cite{COMET:2025sdw}, respectively. The lower-panel shows the predictions for $\mu \to e$ in Gold. The blue shaded region is excluded by the SINDRUM II experiment~\cite{SINDRUMII:2006dvw}.}
    \label{mutoe-Al-Au}
\end{figure}
\begin{figure}[t]
        \includegraphics[width=0.4\textwidth]{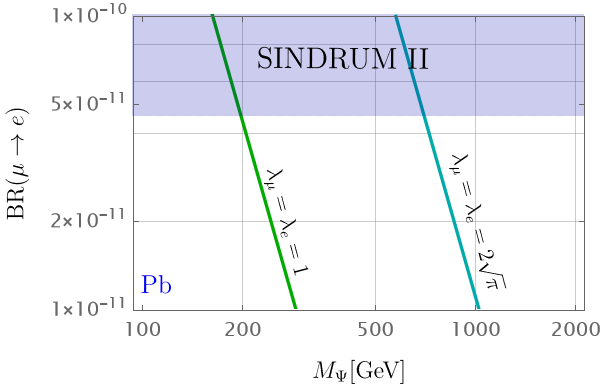}
        \includegraphics[width=0.4\textwidth]{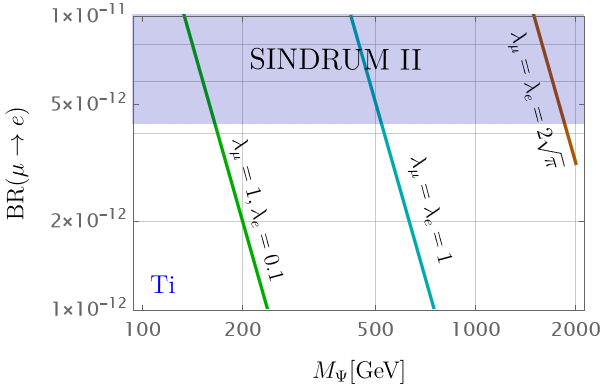}    
    \caption{The upper-panel shows the predictions for $\mu \to e$ conversion in Lead. The blue shaded region is excluded by the SINDRUM II experiment~\cite{SINDRUMII:1996fti}, while in the lower-panel we show the predictions in Titanium. In all scenarios we assume $M_\psi= 2 M_\phi$.}
    \label{fig:mutoeLeadTi}
 \end{figure} 
 
In this theory, the global flavour symmetries, $U(1)_{\ell_e}, U(1)_{\ell_\mu}$ and $U(1)_{\ell_\tau}$, are broken by the interactions in Eq.(\ref{LFV-Yukawa}). In this article, we will investigate the predictions for the most important lepton number violating processes.
These processes occur at one-loop level where inside the loop we have a new electrically charged fermionic field and a neutral scalar field present in the theory. Notice that this theory has only five relevant free parameters for the LFV discussion:
\begin{displaymath}
\lambda_e, \lambda_\mu, \lambda_\tau, M_\Psi, \ {\rm{and}} \ M_\phi.    
\end{displaymath}
and as we discussed above we have an upper bound on the symmetry breaking scale coming from the dark matter relic density constraints.

\begin{itemize}
    \item ${\mathbf{\mu \to e \gamma}}$:
The Feynman diagram for the $\mu \to e \gamma$ is shown in Fig \ref{mueg}. 
The decay width for this process reads as
\begin{equation}
    \Gamma(\mu \to e \gamma)= \frac{e^2 \lambda_e^2 \lambda_\mu^2}{16 \pi}m_\mu^5 |A_L|^2,
\end{equation}
where the $A_L$ coefficient is given in Eq.(\ref{AL}). We provide the full calculation in Appendix~\ref{app:mu2egamma}. 
In Fig.~\ref{fig:mutoegamma} we show the predictions for ${\rm{BR}}(\mu \to e \gamma)$ for different values of the free parameters $\lambda_e$, $\lambda_\mu$, $M_\phi$ and $M_\Psi$. In the upper-panel one can see that if the couplings are between 0.1 and 1, the experimental bounds exclude all scenarios with masses below the TeV scale. 
The dashed red line shows the projected MEG II experiment bound~\cite{Baldini_2021}.
In the lower-panel one can see that if $\lambda_e=\lambda_\mu=1$ one excludes masses up to approximately 30 TeV. In the previous section we mentioned the fact that the relic density constraints studied in Ref.~\cite{Debnath:2025rbu} tell us that one has an upper bound on the symmetry breaking scale around 30 TeV. Therefore, one can see here a possible strong correlation between the cosmological and LFV bounds. These bounds are also very important when we think about the testability of this theory at colliders since the decays of the fermionic field $\Psi$ are determined by the $\lambda_{e_i}$ couplings.

\item ${\mathbf{\tau \to e (\mu)\gamma}}$: In Fig.~\ref{fig:tautoegamma} we show the numerical predictions for the $\tau$ radiative decays. In the upper-panel we show the results for ${\rm{BR}}(\tau \to e \gamma)$, while in the lower-panel the predictions for ${\rm{BR}}(\tau \to \mu \gamma)$ are shown.
In this case the main experimental bounds are from the BABAR collaboration~\cite{Aubert_2010}, excluding the shaded regions in light green. These bounds are weaker than the bounds discussed in Fig.~\ref{fig:mutoegamma} but one can constrain a different combination of Yukawa couplings. Notice that in the scenarios where $\lambda_e,\lambda_\tau \gg \lambda_\mu$ these bounds can be very important.

%%%%%%%%%%%%%%%%%%%%%%%%%%%%%%%%%%%%%%%%
\item {\textit{${\mathbf{\mu \to e \ \rm{conversion}}}$}}:
%%%%%%%%%%%%%%%%%%%%%%%%%%%%%%%%%%%%%% 
\begin{figure}[t]       \includegraphics[width=0.4\textwidth]{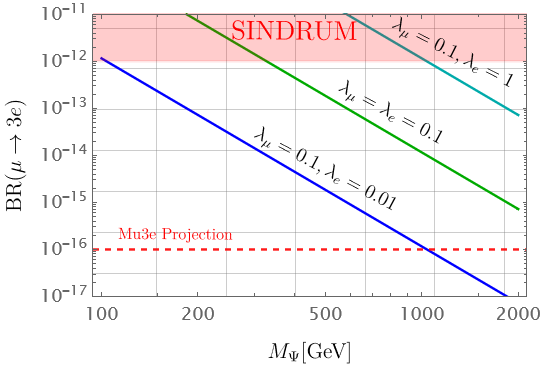}
\includegraphics[width=0.4\textwidth]{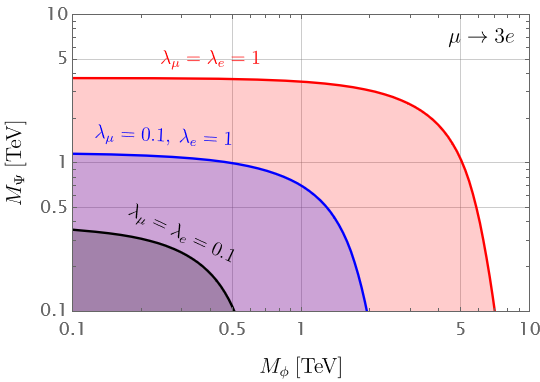}  
\caption{Upper-panel: branching ratio of $\mu \to 3e $ vs. $M_{\Psi}$. The red region is excluded by the SINDRUM bounds~\cite{SINDRUM:1987nra} and the dashed red line shows the projected Mu3e experiment bound~\cite{COMET:2025sdw}. Here we assume $M_\Psi = 2M_\phi$. In the lower-panel one shows the allowed parameter space in the $M_\Psi-M_\phi$ plane for different values of $\lambda_e ,\lambda_\mu$. The shaded regions are excluded by the SINDRUM experiment ~\cite{SINDRUM:1987nra}.}
\label{fig:muonto3e}
 \end{figure}
 %%%%%%%%%%%%%
In the near future, the Mu2e experiment at Fermilab will improve the $\mu \to e$ bounds in four order of magnitude~\cite{COMET:2025sdw}. Therefore, it is important to investigate the predictions for $\mu \to e$ conversion in this theory and show the allowed parameter space.

In Fig.~\ref{mutoe1} we show the photon contribution to $\mu \to e$ conversion in this model, while in Appendix~\ref{App:mu2e} we provide the full set of Feynman graphs. The most important contribution is mediated by the photon. In order to understand the contribution to $\mu \to e$ conversion mediated by the photon we write the general effective LFV current as
\begin{eqnarray}
J_{e\mu}^\nu &=& e \lambda_e \lambda_\mu 
 \bar{e} ( (q^2 \gamma^\nu - \slashed{q} q^\nu) (F_L^{(\gamma)} P_L + F_R^{(\gamma)} P_R) \nonumber \\
&+& i m_\mu \sigma^{\nu \lambda} q_ {\lambda} (A_L P_L + A_R P_R)) \mu.   
\label{jemu}
\end{eqnarray}
Here $q^\nu$ is the photon momentum, $\sigma^{\nu \lambda}=i[\gamma^\nu,\gamma^\lambda]/2$, and $\gamma^\nu$ are the gamma matrices.
In this theory the coefficients $A_R=0$ and $F_L^{(\gamma)}=0$. The $A_L$ coefficient is given in Eq.(\ref{AL}), while $F_R^{(\gamma)}$ is given in Eq.(\ref{FR}). 

The effective Lagrangian for $\mu \to e$ conversion at the quark level can be written as~\cite{Kuno:1999jp} 
\begin{eqnarray}
   - \mathcal{L}_{eff}^{\mu \to e} &=&  m_\mu C_D \ \bar{e}\sigma^{\alpha \beta} P_L \mu F_{\alpha \beta} \nonumber \\
   &+& \sum_{q}C_V^q \ \bar{e} P_L \gamma^\alpha \mu (\bar{q}\gamma_\alpha q) + {\rm{H.c.}}
\end{eqnarray}
Here $C_D= \lambda_e^* \lambda_\mu e A_L/2$ and $C_V^q=\lambda_e^* \lambda_\mu e\hspace{0.1 em} Q^q F_R^{\gamma}$, with $Q^q$ being the electric charge of a given quark. Then, the muon conversion rate can be written as~\cite{Cirigliano:2009bz}
\begin{equation}
    \Gamma_{\mu\to e}= \frac{m_\mu^5}{4} \Bigg|C_D\hspace{0.1 cm} D + \tilde{C}_V^{p} 4V^p +\tilde{C}_V^{n} 4V^n \Bigg|^2,
\end{equation}
while the branching ratio for $\mu $ to e conversion reads as
\begin{eqnarray}
    \text{BR}_{\mu \to e } (Z)= \frac{\Gamma_{\mu\to e} (A,Z)}{\Gamma_{capt}^\mu(A,Z)}.
\end{eqnarray}
See Appendix~\ref{App:mu2e} for a detailed discussion and the values of the input parameters $D$, $\tilde{C}_V^p$, $\tilde{C}_V^n$, $V^p$ and $V^n$.
%%%%%%%%%%%%%%%%%%%
\begin{figure}[h]
\includegraphics[width=0.45\textwidth]{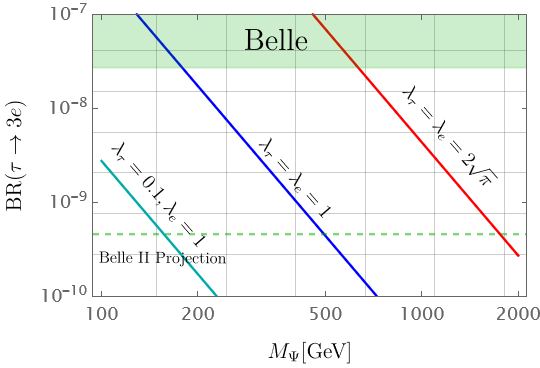}
\includegraphics[width=0.45\textwidth]{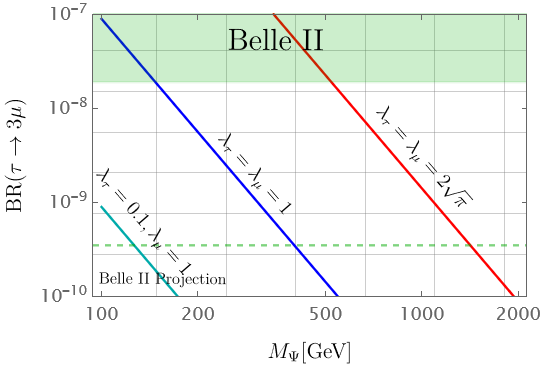} 
\caption{Upper-panel: branching ratio of $\tau \to 3e $ vs. $M_{\Psi}$. The green region is excluded by the Belle experiment~\cite{Hayasaka_2010}, and the dashed green line shows the projected Belle II experiment bound ~\cite{Belle-II:2018jsg}.  In the lower panel, we show the branching ratio of $\tau \to 3 \mu$ as a function of $\Psi$ mass.The green region is excluded by the Belle II experiment ~\cite{2024}, and the dashed green line shows the projected Belle II experiment bound ~\cite{Belle-II:2018jsg}. Here we assume $M_\Psi = 2M_\phi$.}
\label{fig:tauto3e}
 \end{figure}
%%%%%%%%%%%%%%%%%%%

In Fig.~\ref{mutoe-Al-Au} we show the numerical predictions for $\mu \to e$ conversion when we use the Al nucleus (upper-panel), and for Au (lower-panel). In the case of $\mu \rm{Al} \to e \rm{Al}$ we do not have any current bound, but the COMET and Mu2e collaborations will be able to set strong bounds and exclude a large fraction of the parameter space in case of no discovery.
In the lower-panel we show the shaded red region excluded by the SINDRUM II collaboration~\cite{SINDRUMII:2006dvw}. Clearly, the bounds coming from $\mu \to e \gamma$ are stronger but these bounds are very important as an alternative test. 
In Fig.~\ref{fig:mutoeLeadTi} we show the predictions for $\mu \to e$ conversion in Lead and Titanium. In both cases one has experimental bounds from the SINDRUM II collaboration excluding a large fraction of the parameter space.
In this case when the couplings are close to one, the new fields can be below the TeV scale.

\item ${\bf{e_i^\pm \to e_j^\pm e_k^\pm e^\mp_k}}$:
The three-body decays of charged leptons such as $\mu \to 3 e$ are also very unique. In this case one can have contributions mediated by the photon, the gauge bosons, $Z$ and $Z_\ell$, and we can have box-type graphs where inside the loop one has the new fermions and scalar fields. See appendix~\ref{Appe3e} for all Feynman graphs.
In Fig.~\ref{fig:muonto3e} we show the predictions for $\mu \to 3e$ taking into account the main contributions mediated by the photon. In the upper-panel one can see the excluded region by the current experimental bounds and when for example when $\lambda_e=\lambda_\mu=0.1$ one can exclude the region when $M_\Psi < 300$ GeV. The projected bounds from the Mu3e collaboration can test or rule out the region of the parameter space with masses below the TeV scale. In the lower-panel we show the excluded region in $M_{\Psi}-M_\phi$ plane for different scenarios. Notice that when the couplings are equal to one, one excludes masses below 4 TeV. In Fig.~\ref{fig:tauto3e} we show the results for $\tau \to 3 e$ (upper-panel) and $\tau \to 3 \mu$ (lower-panel). The experimental bounds for these processes are not so strong, but they can play a role if the projected bounds are achieved in the near future. 
%%%%%%%%%%%%%%%%%%%%%%%%%%%%%%%%%%%%%%
\item ${\bf{g-2}}$:
%%%%%%%%%%%%%%%%%%%%%%%%%%%%%%%
In this theory the contribution to the muon $g-2$ factor is chirality suppressed. However, we can use the experimental bounds to set an independent bound on the relevant LFV coupling.
Fig.~\ref{fig:muong2} shows the predictions for muon magnetic moment. This contribution can be written as 
\begin{equation}
     a_\mu =2    (A+C+E) \hspace{0.1 em} \lambda_\mu^2 \hspace{0.1 em} m_\mu^2,
\end{equation}
where the explicit formulas for A, C, and E are defined in the appendix~\ref{app:mu2egamma}. Comparing the current experimental average from run $1-6$  of E989 and E821 experiments at Fermilab, one gets~\cite{aliberti2025anomalousmagneticmomentmuon,collaboration2025measurementpositivemuonanomalous}
\begin{eqnarray}
    \Delta a_\mu = a_\mu^{SM}-a_\mu^{exp} = 38 \times 10^{-11}.
\end{eqnarray}
\begin{figure}[t]
        \includegraphics[width=0.45\textwidth]{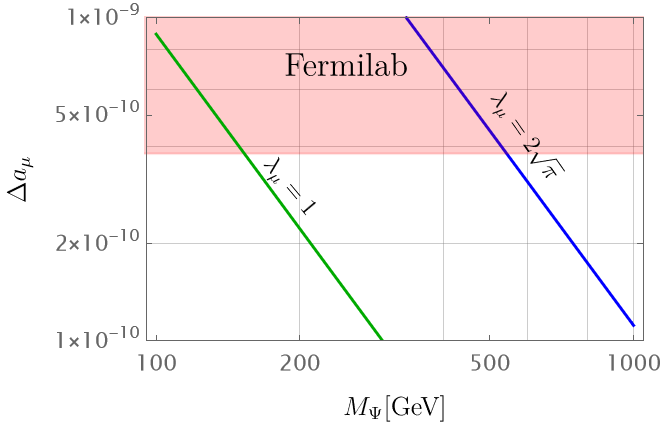}  
\caption{Contribution to the muon magnetic moment vs. $M_\Psi$. Here we assume $M_\Psi=2 M_\phi$.}
\label{fig:muong2}
 \end{figure}
Notice that in this case when the relevant Yukawa coupling is order one, one can rule out the parameter space with masses below 200 GeV. This bound is not very strong but one can constraint $\lambda_\mu$ independently of the other Yukawa couplings. 
\begin{figure}[b]
\includegraphics[width=0.45\textwidth]{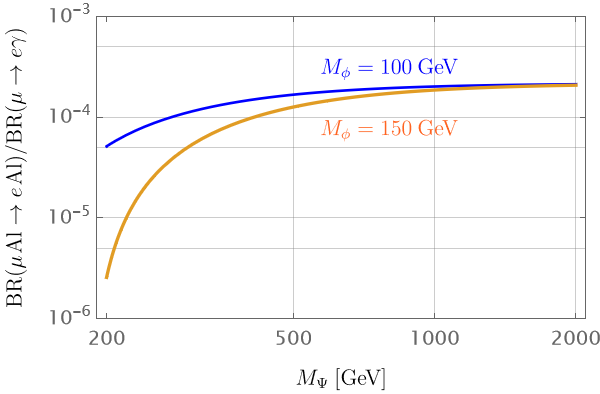}  
\caption{Ratio between branching ratios of  $\mu \to e \gamma$ and $\mu \to e $ conversion as a function of $\Psi$ mass for different values of $\phi $ mass.}
\label{fig:ratio}
\end{figure}

In order to compare the impact of the different processes, we show in Fig.~\ref{fig:ratio} the ratio between the branching ratio for $\mu \to e$ conversion in Aluminium and $\mu \to e \gamma$. We show two examples where the mass of the new scalar field is $100$ GeV (blue line) or $150$ GeV (orange line). 
As one can appreciate, the branching ratio for 
$\mu \to e \gamma$ is $10^{4}$ larger than ${\rm{BR}}(\mu \ \rm{Al} \to e \ \rm{Al})$ in most of the parameter space. 
As a summary, we can list the ratio between the different process where the same combinations of LFV coupling enter:
\begin{eqnarray}
 {\rm{BR}(\mu \ Al \to e \ Al)} &\approx& 10^{-4} \times {\rm{BR}(\mu \to e \gamma)}, \\
  {\rm{BR}(\mu \to 3 e )} &\approx& 10^{-3} \times {\rm{BR}(\mu \to e \gamma)}, \\
  {\rm{BR}(\tau \to 3 \mu )} &\approx&  10^{-3} \times {\rm{BR}(\tau \to \mu \gamma)},\\
  {\rm{BR}(\tau \to 3 e )} &\approx& 10^{-2} \times {\rm{BR}(\tau \to e \gamma)}.
\end{eqnarray}
Clearly, these results tell us that in order to test this theory for neutrino masses one needs to find the radiative decays of the muon and tau leptons, and cross-check the ratios listed above.
%%%%%%%%%%%%%%%%%%%%%%%%%%%%%%%%%%%%%%%%%%%

\begin{figure}[t]
        \includegraphics[width=0.45\textwidth]{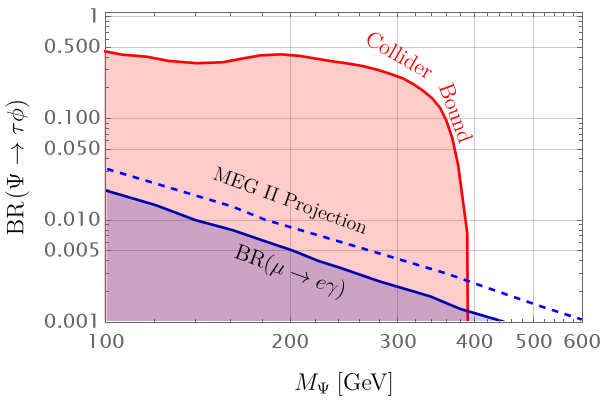}  
\caption{Correlation between the collider and lepton flavor violating bounds. The red shaded region is excluded from the collider study in Ref.~\cite{Butterworth:2025asm}, while the blue region is excluded by $\mu \to e \gamma$ constraints. For illustration, we assume $\lambda_e=\lambda_\mu$, and scan all $\lambda_i$ from 0.001 to 0.1 assuming $M_\phi=50$ GeV.}
\label{fig:collider}
 \end{figure}

%%%%%%%%%%%%%%%%%%%%%%%%%%%%%%%%%%%%%%%%
\item \textit{Collider vs. LFV Constraints}:
%%%%%%%%%%%%%%%%%%%%%%%%%%%%%%%%%%%%%%%%
The theory investigated in this article is quite simple and one can hope to test the predictions at current and future colliders. In section~\ref{theory} we discussed the main signatures at colliders, we mentioned the signatures with charged tracks and the channels with mutileptons and missing energy. In Ref.~\cite{Butterworth:2025asm} we studied the collider signatures and pointed out the lower bounds on the fermion masses using the channels with multi-leptons and missing energy. We have shown that one can ruled out several scenarios where the branching ratio $\rm{BR}(\Psi \to e \phi)$ or $\rm{BR}(\Psi \to \mu \phi)$ are large and the fermion masses are below the TeV scale. In the scenarios when $\rm{BR}(\Psi \to \tau \phi)$ is large one can use the searches with di-taus and missing energy to exclude the region of the parameter space when the $M_\Psi$ is below 400 GeV. For a detailed discussion see Ref.~\cite{Butterworth:2025asm}. 

In order to show the impact of the collider bounds, we show in Fig.~\ref{fig:collider} the correlation between the collider and lepton flavour violating bounds. In this case the 
$\rm{BR}(\Psi \to \tau \phi)$ is large. The red shaded region is excluded from the collider bounds found in Ref.~\cite{Butterworth:2025asm}, while the blue region is excluded by $\mu \to e \gamma$ searches. For illustration, we have assumed $\lambda_e=\lambda_\mu$, and scan all $\lambda_i$ from 0.001 to 0.1 assuming $M_\phi=50$ GeV. One can appreciate that in these scenarios the collider bounds are much stronger, ruling out most of the parameter space when the $\Psi$ mass is below 400 GeV, while the $\mu \to e \gamma$ bounds are not so important. Of course, as we have discussed above, when the Yukawa couplings, $\lambda_e$ and $\lambda_\mu$, are large the $\mu \to e \gamma$ bounds are more important.

\end{itemize}

%%%%%%%%%%%%%%%%%%%%%%%
\section{SUMMARY}
\label{summary]}
%%%%%%%%%%%%%%%%%%%%%%%
It is well established that the observation of lepton number violating processes in the charged lepton sector of the Standard Model would constitute unambiguous evidence for the existence of physics beyond the Standard Model. Owing to the remarkable progress of several experimental collaborations, such as Mu2e, the detection of such processes may become feasible in the near future.
In this work, we examine the predictions for lepton flavor violating processes within the minimal theory of neutrino masses, which is formulated on the basis of the spontaneous breaking of local lepton number at the low energy scale. This theory predicts the existence of four additional fermionic states required by gauge anomaly cancellation, one of which emerges as a viable dark matter candidate. These new fermions acquire their masses from the same symmetry breaking mechanism, whose scale is bounded from above by the requirement of reproducing the observed dark matter relic abundance. Given that this theory can describe phenomena below the multi-TeV scale, it becomes imperative to investigate its implications for low-energy observables with high precision.

Within this theory, LFV processes arise at the one-loop level through diagrams involving the new charged fermions and a neutral scalar field. Thanks to the simplicity of this theory, its phenomenological consequences can be comprehensively tested through a combined analysis of LFV searches and collider experiments. We presented a detailed study of the relevant LFV processes and, by employing current experimental bounds, delineate the corresponding allowed regions in parameter space. Furthermore, by incorporating projected sensitivities, we identify the ranges of Yukawa couplings and new particle masses that can be probed in forthcoming experimental efforts. This minimal theory of neutrino masses thus provides a coherent and testable framework that simultaneously accounts for the origin of neutrino masses, lepton number violation, and dark matter, offering a concrete avenue toward uncovering the structure of new physics at accessible energy scales.
%
%%%%%%%%%%%%%%%%%%%%%%%%%%%%%%%%%%%%%%%%%
\appendix
\begin{widetext}
%%%%%%%%%%%%%%%%%%%%%%%%%%%%%%%%%%%%%%%%
\section{Experimental Bounds}
%%%%%%%%%%%%%%%%%%%%%%%%%%%%%%%%%%%%%%%
\begin{table}[h]
    \centering
\begin{tabular}{|c|c|c|}
\hline
\text{\makecell{LFV \\ Process }} & \makecell{Current \\ Upper Limit} & \makecell{Future  \\ Projection }  \\
\hline
$\mu \to e \gamma$ & $ 1.5 \times 10^{-13}$~\cite{afanaciev2025newlimitmuegammadecay} & $6.0\times 10^{-14}$ ~\cite{Baldini_2021}  \\
\hline
$\tau \to e \gamma$ & $  3.3 \times 10^{-8}$ ~\cite{Aubert_2010}& $3.0 \times 10 ^{-9}$~\cite{Belle-II:2018jsg}  \\
\hline
$\tau \to \mu \gamma$ &  $  4.2 \times 10^{-8}$ ~\cite{Belle:2021ysv}& $1.0 \times 10^{-9}$~\cite{Belle-II:2018jsg}   \\
\hline
$\mu \to 3 e$ & $ 1.0\times 10^{-12}$~\cite{SINDRUM:1987nra} & $ \sim 10^{-16}$~\cite{COMET:2025sdw} \\
\hline
$\tau \to 3 e$ & $ 2.7\times 10^{-8} $~\cite{Hayasaka_2010} & $   4.6 \times 10^{-10}$ ~\cite{Belle-II:2018jsg} \\
\hline
$\tau \to 3 \mu$ & $ 1.9\times 10^{-8} ~\cite{2024}$ & $   3.5 \times 10^{-10}$ ~\cite{Belle-II:2018jsg} \\
\hline
$\mu\hspace{0.1 cm} \text{Al} \to  e \hspace{0.1 cm} \text{Al}$ & - & $ \sim 10^{-17}$ ~\cite{COMET:2025sdw}\\
\hline
$\mu\hspace{0.1 cm} \text{Ti} \to  e \hspace{0.1 cm} \text{Ti}$ & $ 4.3 \times 10^{-12}$ ~\cite{SINDRUMII:1993gxf}& - \\
\hline
$\mu\hspace{0.1 cm} \text{Au} \to  e \hspace{0.1 cm} \text{Au}$ & $ 7.0 \times 10^{-13}$~\cite{SINDRUMII:2006dvw} & - \\
\hline
$\mu\hspace{0.1 cm} \text{Pb} \to  e \hspace{0.1 cm} \text{Pb}$ & $ 4.6 \times 10^{-11}$~\cite{SINDRUMII:1996fti} & - \\
\hline
\end{tabular}
\caption{LFV bounds.}
\label{ExpBounds}
\end{table}
%%%%%%%%%%%%%%%%%%%%%%%%%%%%%%%%%%%%%%%%
\section{$\mu \to e \gamma$}
\label{app:mu2egamma}
%%%%%%%%%%%%%%%%%%%%%%%%%%%%%%%%%%%%%%%
The relevant Lagrangian to discuss the lepton number violating processes is given by
\begin{equation*}
    \mathcal{L}  \supset  \lambda_{e_i} \bar{\Psi}_L e_R^i \phi + {\rm{H.c.}} 
\end{equation*}
The one-loop  diagram for $\mu (p_1) \to e (p_2) \hspace{ 0.1 cm}\gamma(q)$   is shown in Fig.~\ref{mueg}, where $ p_2 = p_1-q .$
The amplitude for this process reads as 
\begin{align}
    i \mathcal{M}_1 & =   \bar{u}_e (p_2)\int \frac{d^4 k}{(2 \pi)^4} i\lambda_e P_L \frac{i (\slashed{k} +\slashed{p_2} + M_\Psi)}{(k+p_2)^2 -M_\Psi^2}(- i e \gamma^\nu) \frac{i (\slashed{k} + \slashed{p_1} + M_\Psi)}{(k+p_1)^2 -M_\Psi^2} i\lambda_\mu P_R \frac{i}{k^2 -M_\phi^2 } u_\mu(p_1) \epsilon_\nu(q) \nonumber \\
    &= e \lambda_e \lambda_\mu \bar{u}_e(p_2) \int\frac{d^4k}{(2 \pi)^4} \frac{P_L (\slashed{k}+\slashed{p_2}+ M_\Psi) \gamma^\nu (\slashed{k} +\slashed{p_1}+ M_\Psi) P_R}{[(k+p_1)^2 -M_\Psi^2] [(k+p_2)^2 -M_\Psi^2][k^2 -M_\phi^2]} u_\mu(p) \epsilon_{\nu}(q).
\end{align}
One can simplify the numerator as follows:
\begin{align*}
    N_\nu & = P_L (\slashed{k}+\slashed{p_2}+ M_\Psi) \gamma^\nu (\slashed{k} +\slashed{p_1}+ M_\Psi) P_R 
     = P_L( \slashed{k}\gamma^\nu \slashed{k}+\slashed{k}\gamma^\nu \slashed{p_1}+\slashed{p_2}\gamma^\nu \slashed{k}+\slashed{p_2}\gamma^\nu \slashed{p_1}+M_\Psi^2 \gamma^\nu ).
\end{align*}
Notice that any term proportional to $\bar{u}_e(p_2) \gamma^\nu u_{\mu}(p_1)$ will cancel out with the contribution from the diagram when the photon is attached to either $\mu$ or $e$ external line. 
The loop functions can be written as
\begin{align}
     \int\frac{d^4k}{(2 \pi)^4} \frac{k^\alpha k^\beta}{[(k+p_1)^2 -M_\Psi^2] [(k+p_2)^2 -M_\Psi^2][k^2 -M_\phi^2]} 
    & = A p_1^\alpha p_1^\beta+ B p_2^\alpha p_2^\beta + C( p_1^\alpha p_2^\beta+ p_2^\alpha p_1^\beta) + D g^{\alpha\beta}, \\
    \int\frac{d^4k}{(2 \pi)^4} \frac{k^\alpha }{[(k+p_1)^2 -M_\Psi^2] [(k+p_2)^2 -M_\Psi^2][k^2 -M_\phi^2]} & =
    E p_1^\alpha + F p_2^\alpha, \\
    \int\frac{d^4k}{(2 \pi)^4} \frac{1 }{[(k+p_1)^2 -M_\Psi^2] [(k+p_2)^2 -M_\Psi^2][k^2 -M_\phi^2]} &= C_0. 
\end{align}
where $A, B, C, D, E, F$ are some parameters that don't depend on the momentum.  Performing the integration, we get
\begin{align*}
    \bar{u}_e(p_2) P_L \mathcal{N}^\nu u_\mu(p_1)& = \bar{u}_e(p_2) P_L (A \hspace{0.1 cm} \slashed{p_1}\gamma^\nu \slashed{p_1}+ B \hspace{0.1 cm} \slashed{p_2}\gamma^\nu \slashed{p_2} + C \hspace{0.1 cm} \slashed{p_1}\gamma^\nu \slashed{p_2}+ C \hspace{0.1 cm}\slashed{p_2}\gamma^\nu \slashed{p_1} \\
    & + E\hspace{0.1 cm} \slashed{p_1}\gamma^\nu \slashed{p_1}+(E+F)\hspace{0.1 cm} \slashed{p_2}\gamma^\nu \slashed{p_1} +F \hspace{0.1 cm} \slashed{p_2}\gamma^\nu \slashed{p_2}+ \slashed{p_2}\gamma^\nu \slashed{p_1} + C_0 M_\Psi^2 \gamma^\nu+D g^{\alpha \beta} \gamma^\alpha \gamma^\nu \gamma^\beta)u_\mu(p_1),
\end{align*}
where the divergent D term gets cancelled out with the contribution from the other diagrams mentioned above. One can  simplify further by the applying the Gordon's identity and we write 
\begin{align}
   \hspace{0.1 cm}\bar{u}_e(p_2) P_L\slashed{p_1}\gamma^\nu \slashed{p_1} u_\mu(p_1) & =i  m_\mu  \hspace{0.1 cm} \bar{u}_e(p_2) \sigma^{\nu \alpha} q_\alpha P_L u_\mu (p_1) +  m_\mu  \bar{u}_e(p_2) P_L u_\mu(p_1)q^\nu +m_\mu m_e\bar{u}_e(p_2)\gamma^\nu P_L u_\mu(p_1),  \\
     \hspace{0.1 cm}
     \bar{u}_e(p_2) P_L\slashed{p_1}\gamma^\nu \slashed{p_2} u_\mu(p_1) & = i  m_\mu  \hspace{0.1 cm} \bar{u}_e(p_2) \sigma^{\nu \alpha} q_\alpha P_L u_\mu (p_1) + m_\mu  \bar{u}_e(p_2) P_L u_\mu(p_1)q^\nu +m_e m_\mu \bar{u}_e(p_2)\gamma^\nu P_L u_\mu(p_1) \nonumber \\
     &  + i m_e \hspace{0.1 cm} \bar{u}_e(p_2) \sigma^{\nu \alpha} q_\alpha P_R u_\mu (p_1) - m_e \bar{u}_e(p_2) P_R u_\mu(p_1)q^\nu - 2 \bar{u}_e(p_2) \slashed{q} P_R u_\mu(p_1)q^\nu, \\
      \hspace{0.1 cm}\bar{u}_e(p_2) P_L\slashed{p_2}\gamma^\nu \slashed{p_1} u_\mu(p_1) & =m_\mu m_e\bar{u}_e(p_2)\gamma^\nu P_L u_\mu(p_1), \\
     \hspace{0.1 cm}\bar{u}_e(p_2)P_L \slashed{p_2}\gamma^\nu \slashed{p_2} u_\mu(p_1) & =m_e\left(m_\mu \bar{u}_e(p_2)\gamma^\mu P_L u_\mu(p_1) + i \bar{u}_e(p_2) \sigma^{\nu \alpha} q_\alpha P_R u_\mu (p_1) - \bar{u}_e(p_2) P_R u_\mu(p_1)q^\nu  \right). 
\end{align}

In the limit $m_e \to 0$ equations B7 and B8 are zero. One can write the amplitude as  
\begin{equation}
    i \mathcal{M}_1 = i e \lambda_e \lambda_\mu A_L m_\mu  \hspace{0.1 cm} \bar{u}_e(p_2) \sigma^{\nu \alpha} q_\alpha P_L u_\mu (p_1)  \epsilon_{\nu}(q) + (C_0 M_\Psi^2 + D (2-d)) e \lambda_e \lambda_\mu  \hspace{0.1 cm} \bar{u}_e(p_2)P_L \gamma^\nu u_{\mu}(p_1)\epsilon_\nu(q),
\end{equation}
where $A_L= A+C+E$ and spacetime dimension $d= 4 - 2 \epsilon$. For the sake of completeness of the discussion, here we show the contributions from the other diagrams also. 
When the photon is attached to the outgoing electron, the amplitude can be written as 
\begin{align}
    i \mathcal{M}_2 & = e \lambda_e \lambda_\mu \bar{u}_e(p_2)\int \frac{d^4k}{(2\pi)^4} \frac{\gamma^\nu (\slashed{p_1}+m_e) P_L (\slashed{k}+\slashed{p_1}+M_\Psi)P_R}{[(k+p_1)^2-M_\Psi^2][k^2-M_\phi^2][p_1^2-m_e^2]} u_\mu(p_1) \epsilon_\nu(q).
\end{align} 
Now, we write  
\begin{align}
    \int \frac{d^4k}{(2\pi)^4}\frac{k^\alpha}{[(k+p_1)^2-M_\Psi^2][k^2-M_\phi^2]} & = B_1 p_1^\alpha ,\\
      \int \frac{d^4k}{(2\pi)^4}\frac{1}{[(k+p_1)^2-M_\Psi^2][k^2-M_\phi^2]} & = B_0.
\end{align}
Following the same steps mentioned above, 
in the limit $m_e \to 0$, one can write the amplitude as 
\begin{equation}
    i \mathcal{M}_2= e \lambda_e \lambda_\mu(B_1+B_0) \bar{u}_e(p_2)P_L \gamma^\nu u_\mu(p_1) \epsilon_\nu(q).
\end{equation}
Finally, when the photon is attached to the muon, the amplitude can be written as
\begin{equation}
    i \mathcal{M}_3 = e \lambda_e \lambda_\mu \bar{u}_e(p_2) \int \frac{P_L (\slashed{k}+\slashed{p_2}+M_\Psi)P_R (\slashed{p_2}+m_\mu) \gamma^\nu}{[(k+p_2)^2-M_\Psi^2][k^2-M_\phi^2][p_2^2-m_\mu^2]} u_\mu(p_1)\epsilon_\nu(q).
\end{equation}
This contribution is always proportional to the electron mass and in the limit $m_e \to 0$, the contribution vanishes. Therefore, the total amplitude can be written as 
\begin{eqnarray}
    i \mathcal{M}&=&i \mathcal{M}_1+i \mathcal{M}_2, \nonumber \\
    &=&  i e \lambda_e \lambda_\mu A_L m_\mu  \hspace{0.1 cm} \bar{u}_e(p_2) \sigma^{\nu \alpha} q_\alpha P_L u_\mu (p_1)  \epsilon_{\nu}(q) + (C_0 M_\Psi^2 + D (2-d) +B_1 +B_0) e \lambda_e \lambda_\mu  \hspace{0.1 cm} \bar{u}_e(p_2)P_L \gamma^\nu u_{\mu}(p_1)\epsilon_\nu(q). \nonumber \\
\end{eqnarray}
Now, one can show that the coefficient $(C_0 M_\Psi^2 + D (2-d) +B_1 +B_0)=0$ and the amplitude  becomes
\begin{equation}
   i \mathcal{M} =  i e \lambda_e \lambda_\mu A_L m_\mu  \hspace{0.1 cm} \bar{u}_e(p_2) \sigma^{\nu \alpha} q_\alpha P_L u_\mu (p_1)  \epsilon_{\nu}(q). 
\end{equation}
When $M_\Psi,M_\phi >> m_\mu$, one finds
\begin{equation}
    A_L = \frac{i}{16 \pi^2}\left( \frac{2M_\phi^6+3M_\phi^4M_\Psi^2-6 M_\phi^2M_\Psi^4+M_\Psi^6-6M_\phi^4 M_\Psi^2 \ln\left(M_\phi^2/M_\Psi^2\right)}{12 (M_\phi^2-M_\Psi^2)^4} \right).
    \label{AL}
\end{equation}
We use Package-X~\cite{Patel:2015tea,Patel:2016fam} to perform all loop calculations.
%%%%%%%%%%%%%%%%%%%%%%%%%%%%%%%%%%%%%%%
\section{$\mu \to e$ conversion}
\label{App:mu2e}
%%%%%%%%%%%%%%%%%%%%%%%%%%%%%%%%%%%%%%%
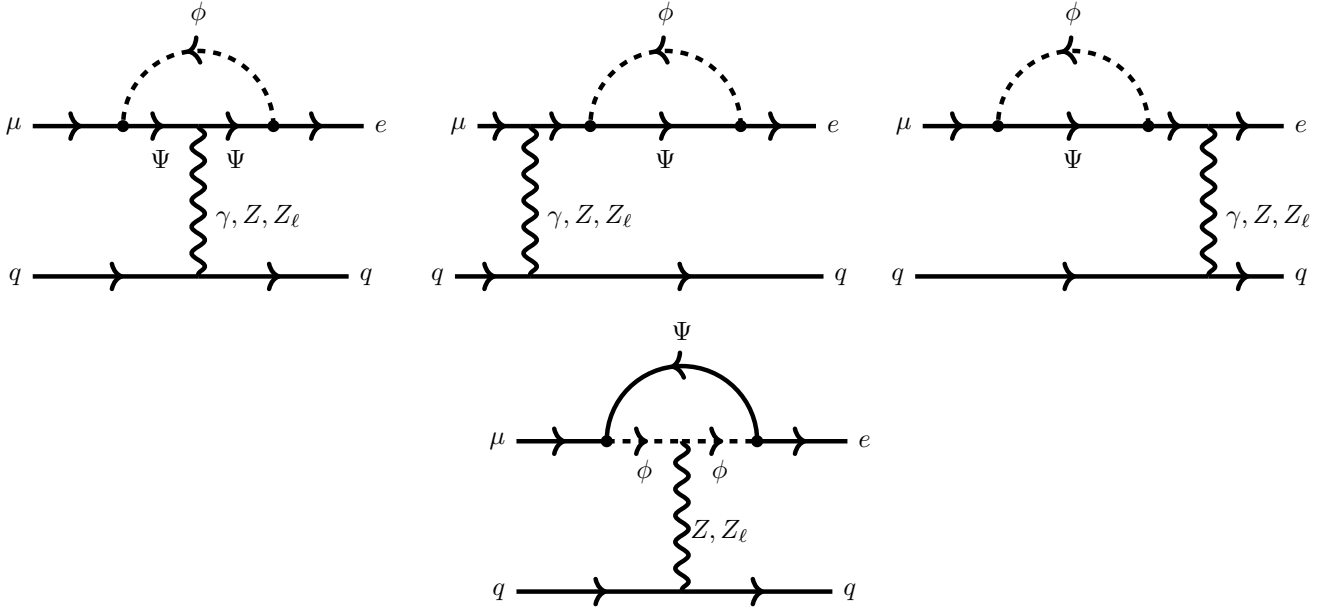
\begin{figure}[h]
\begin{eqnarray*}
\begin{gathered}
\begin{tikzpicture}[line width= 1.7 pt,node distance=1 cm and 1 cm]
\coordinate[label=left:$\mu$](a);
\coordinate[right = 1.2 cm of a](v1);
\coordinate[right = 2 cm of v1](v2);
\coordinate[right = 1 cm of v1](aux);
\coordinate[above = 1.2 cm of aux,label=$\phi$](phi);
\coordinate[right = 1.2 cm of v2,label=right:$e$](e);
\coordinate[right = 1 cm of v1](psi);
\coordinate[below = 2 cm of psi](psi2);
\coordinate[left = 2.2 cm of psi2,label=left:$q$](q2);
\coordinate[right = 2.0 cm of psi2,label=right:$q$](q3);
\coordinate[right = 0.5 cm of v1](psi3);
\coordinate[right = 1.5 cm of v1](psi4);
\coordinate[right = 1.8 cm of v1](g1);
\coordinate[below = 0.7 cm of psi3,label=$\Psi$](psi5);
\coordinate[below = 0.7 cm of psi4,label=$\Psi$](psi6);
\coordinate[below = 1.5 cm of g1,label={$\gamma,Z,Z_\ell$}](gamma);
\draw[fill=black](v1) circle (.05cm);
\draw[fill=black](v2) circle (.05cm);
\draw[fermion](a)--(v1);
\draw[fermion](v1)--(psi);
%\draw[fermionnoarrow](v1)--(v2);
\draw[fermion](psi)--(v2);
\draw[fermion](v2)--(e);
\draw[vector](psi) -- (psi2);
\draw[fermion](q2)--(psi2);
\draw[fermion](psi2)--(q3);
\semiloop[scalar]{v1}{v2}{0};
\end{tikzpicture}
\end{gathered} 
\quad   \! 
\begin{gathered}
\begin{tikzpicture}[line width= 1.7 pt,node distance=1 cm and 1 cm]
\coordinate[label=left:$\mu$](a);
\coordinate[right = 0.7 cm of a](v3);
\coordinate[right = 1.5 cm of a](v1);
\coordinate[right = 2 cm of v1](v2);
\coordinate[right = 1 cm of v1](aux);
\coordinate[above = 1.2 cm of aux,label=$\phi$](phi);
\coordinate[right = 1.0 cm of v2,label=right:$e$](e);
\coordinate[right = 0.51 cm of v1](psi);
\coordinate[below = 2 cm of v3](psi2);
\coordinate[left = 1.0 cm of psi2,label=left:$q$](q2);
\coordinate[right = 3.9 cm of psi2,label=right:$q$](q3);
\coordinate[right = 0.5 cm of v1](psi3);
\coordinate[right = 1.0 cm of v1](psi4);
\coordinate[right =  0.8 cm of v3](g1);
%\coordinate[below = 0.7 cm of psi3,label=$\Psi$](psi5);
\coordinate[below = 0.7 cm of psi4,label=$\Psi$](psi6);
\coordinate[below = 1.5 cm of g1,label={$\gamma,Z,Z_\ell$}](gamma);
\draw[fill=black](v1) circle (.05cm);
\draw[fill=black](v2) circle (.05cm);
\draw[fermion](a)--(v3);
\draw[fermion](v3)--(v1);
\draw[fermion](v1)--(v2);
%\draw[fermionnoarrow](v1)--(v2);
%\draw[fermion](psi)--(v2);
\draw[fermion](v2)--(e);
\draw[vector](v3) -- (psi2);
\draw[fermion](q2)--(psi2);
\draw[fermion](psi2)--(q3);
\semiloop[scalar]{v1}{v2}{0};
\end{tikzpicture}
\end{gathered}
\quad   \! 
\begin{gathered}
\begin{tikzpicture}[line width= 1.7 pt,node distance=1 cm and 1 cm]
\coordinate[label=left:$\mu$](a);
\coordinate[right = 1.0 cm of a](v1);
\coordinate[right = 2 cm of v1](v2);
\coordinate[right = 0.8 cm of v2](v3);
\coordinate[right = 1 cm of v1](aux);
\coordinate[above = 1.2 cm of aux,label=$\phi$](phi);
\coordinate[right = 1.0 cm of v3,label=right:$e$](e);
\coordinate[right = 0.51 cm of v1](psi);
\coordinate[below = 2 cm of v3](psi2);
\coordinate[left = 3.9 cm of psi2,label=left:$q$](q2);
\coordinate[right = 1.0 cm of psi2,label=right:$q$](q3);
\coordinate[right = 0.5 cm of v1](psi3);
\coordinate[right = 1.0 cm of v1](psi4);
\coordinate[right =  0.8 cm of v3](g1);
%\coordinate[below = 0.7 cm of psi3,label=$\Psi$](psi5);
\coordinate[below = 0.7 cm of psi4,label=$\Psi$](psi6);
\coordinate[below = 1.5 cm of g1,label={$\gamma,Z,Z_\ell$}](gamma);
\draw[fill=black](v1) circle (.05cm);
\draw[fill=black](v2) circle (.05cm);
\draw[fermion](a)--(v1);
%\draw[fermion](v3)--(v1);
\draw[fermion](v1)--(v2);
%\draw[fermionnoarrow](v1)--(v2);
%\draw[fermion](psi)--(v2);
\draw[fermion](v2)--(v3);
\draw[fermion](v3)--(e);
\draw[vector](v3) -- (psi2);
\draw[fermion](q2)--(psi2);
\draw[fermion](psi2)--(q3);
\semiloop[scalar]{v1}{v2}{0};
\end{tikzpicture}
\end{gathered}\\
\begin{gathered}
\text{\makebox[\textwidth][c]{%
  \begin{tikzpicture}[line width=1.7pt,node distance=1 cm and 1 cm, baseline=(current bounding box.center)]
   \coordinate[label=left:$\mu$](a);
\coordinate[right = 1.2 cm of a](v1);
\coordinate[right = 2 cm of v1](v2);
\coordinate[right = 1 cm of v1](aux);
\coordinate[above = 1.2 cm of aux,label=$\Psi$](phi);
\coordinate[right = 1.2 cm of v2,label=right:$e$](e);
\coordinate[right = 1 cm of v1](psi);
\coordinate[below = 2 cm of psi](psi2);
\coordinate[left = 2.2 cm of psi2,label=left:$q$](q2);
\coordinate[right = 2.0 cm of psi2,label=right:$q$](q3);
\coordinate[right = 0.5 cm of v1](psi3);
\coordinate[right = 1.5 cm of v1](psi4);
\coordinate[right = 1.5 cm of v1](g1);
\coordinate[below = 0.7 cm of psi3,label=$\phi$](psi5);
\coordinate[below = 0.7 cm of psi4,label=$\phi$](psi6);
\coordinate[below = 1.5 cm of g1,label={$Z,Z_\ell$}](gamma);
\draw[fill=black](v1) circle (.05cm);
\draw[fill=black](v2) circle (.05cm);
\draw[fermion](a)--(v1);
\draw[scalar](v1)--(psi);
%\draw[fermionnoarrow](v1)--(v2);
\draw[scalar](psi)--(v2);
\draw[fermion](v2)--(e);
\draw[vector](psi) -- (psi2);
\draw[fermion](q2)--(psi2);
\draw[fermion](psi2)--(q3);
\semiloop[fermion]{v1}{v2}{0};
  \end{tikzpicture}%
}}
\end{gathered}
\end{eqnarray*}
 \caption{Feynman diagrams for $\mu $ to e conversion. }
    \label{mutoe}
\end{figure}
In Fig.~\ref{mutoe} we show the different contributions to $\mu \to e$ conversion in this model. The dominate contribution is mediated by the photon. Neglecting the lepton masses, the loop factor in Eq.(\ref{jemu}), reads as
\begin{eqnarray}
F_R^{(\gamma)} &=&\frac{i}{16 \pi^2}\left(\frac{11 M_\phi^4 - 7 M_\phi^2 M_\Psi^2 + 2 M_\Psi^4}{36 (M_\phi^2 - M_\Psi^2)^3}
- \frac{M_\phi^6 \ln \left(M_\phi^2/M_\Psi^2 \right)}{6 (M_\phi^2 - M_\Psi^2)^4}\right), \label{FR}
\end{eqnarray}
while $A_R = 0$, and $F_L^{(\gamma)} = 0$.
The $A_L$ coefficient is given in Eq.(\ref{AL}). 
At the low energy scale, one can match between the quark and parton level by introducing form factors, and the matching relation is~\cite{Kitano:2002mt}
\begin{equation}
    \bar{q}\gamma^\nu q \to F_{V_N}^q \bar{\Psi}_N \gamma^\nu \Psi_N,
\end{equation}
where 
\begin{align}
    & F_{V_p}^{(u)} = 2 , \hspace{0.2 cm}
F^{(d)}_{V_p} = 1, \hspace{0.2 cm}
F^{(s)}_{V_p} = 0, \hspace{0.2 cm}
F_{V_n}^{(u)} = 1 , \hspace{0.2 cm}
F^{(d)}_{V_n} = 2, \hspace{0.2 cm} {\rm{and}} \hspace{0.2 cm}
F^{(s)}_{V_n} = 0.
\end{align}
The effective Lagrangian at the nucleon level can be written as 
\begin{align}
   - \mathcal{L}_{eff}^{\mu \to e} = \sum_{N=p,n}  \left[m_\mu C_D \bar{e}\sigma^{\alpha \beta} P_L \mu F_{\alpha \beta} + \left(\tilde{C}_V^N \bar{e} P_L \gamma^\alpha \mu \right)\bar{\Psi}_N\gamma_\alpha \Psi_N + {\rm{H.c.}} \right],
\end{align}
and the new vector couplings are defined as 
\begin{equation}
    \tilde{C}_V^p= \sum_{q=u,d,s} C_V^q F_{V_p}^q \hspace{0.2 cm} \text{and } \hspace{0.2 cm}\tilde{C}_V^n= \sum_{q=u,d,s} C_V^q F_{V_n}^q.
\end{equation}
For convenience, we list the values of the  parameters  used for different nuclei in Table~\ref{tab:CoefficientChart}.

\begin{table}[h]
\centering
\begin{tabular}{|c|c|c|c|c|}
\hline
\text{Nucleus} & \text{D} ~\cite{Kitano:2002mt} & \textbf{$V^p$} ~\cite{Kitano:2002mt} & \textbf{$V^n$} ~\cite{Kitano:2002mt} & \text{$\Gamma_{capt}^\mu (10^{6} s^{-1})$ ~\cite{Suzuki:1987jf}} \\
\hline
Al & 0.0362 & 0.0161 & 0.0173 & 0.69 \\
\hline
Ti & 0.0864 & 0.0396 & 0.0468 & 2.59 \\
\hline
Au & 0.1890 & 0.0974 & 0.1460 & 13.07 \\
\hline
Pb & 0.1610 & 0.0834 & 0.1280 & 13.45 \\
\hline
\end{tabular}
\caption{Values of the overlap integrals for different nuclei. }
\label{tab:CoefficientChart}
\end{table}
 %%%%%%%%%%%%%%%%%%%%%%%%%%%%%%%%%%%%%%%%%%%%%%%%%%%%%%%
\section{$e_i^{\pm} \to e_j^{\pm} e_k^{\pm} e^{\mp}_k$}
\label{Appe3e}
%%%%%%%%%%%%%%%%%%%%%%%%%%%%%%%%%%%%%%%%%%%%%%%%%%%%%%%%
In Fig.~\ref{muto3e}, we show all possible class of Feynman graphs for $e_i \to e_j e_k e_k$ decays at the one-loop level. The Feynman diagrams in the first row of Fig.~\ref{muto3e} that involve $Z$ and $Z_\ell$ are suppressed due to their large mass. When the gauge boson is attached to the scalar $\phi$ , this contribution is suppressed by the small kinetic mixing between $ Z$ and $Z_\ell$. Besides this, $\mu \to e \gamma$ puts a strong bound on the coupling parameter $\lambda_e \hspace{0.1 cm} \text{and}  \hspace{0.1 cm} \lambda_\mu$. The amplitude for the box diagrams is proportional to $\lambda_{e_i}^4$ and therefore is much suppressed compared to the photon contribution to this decay process.
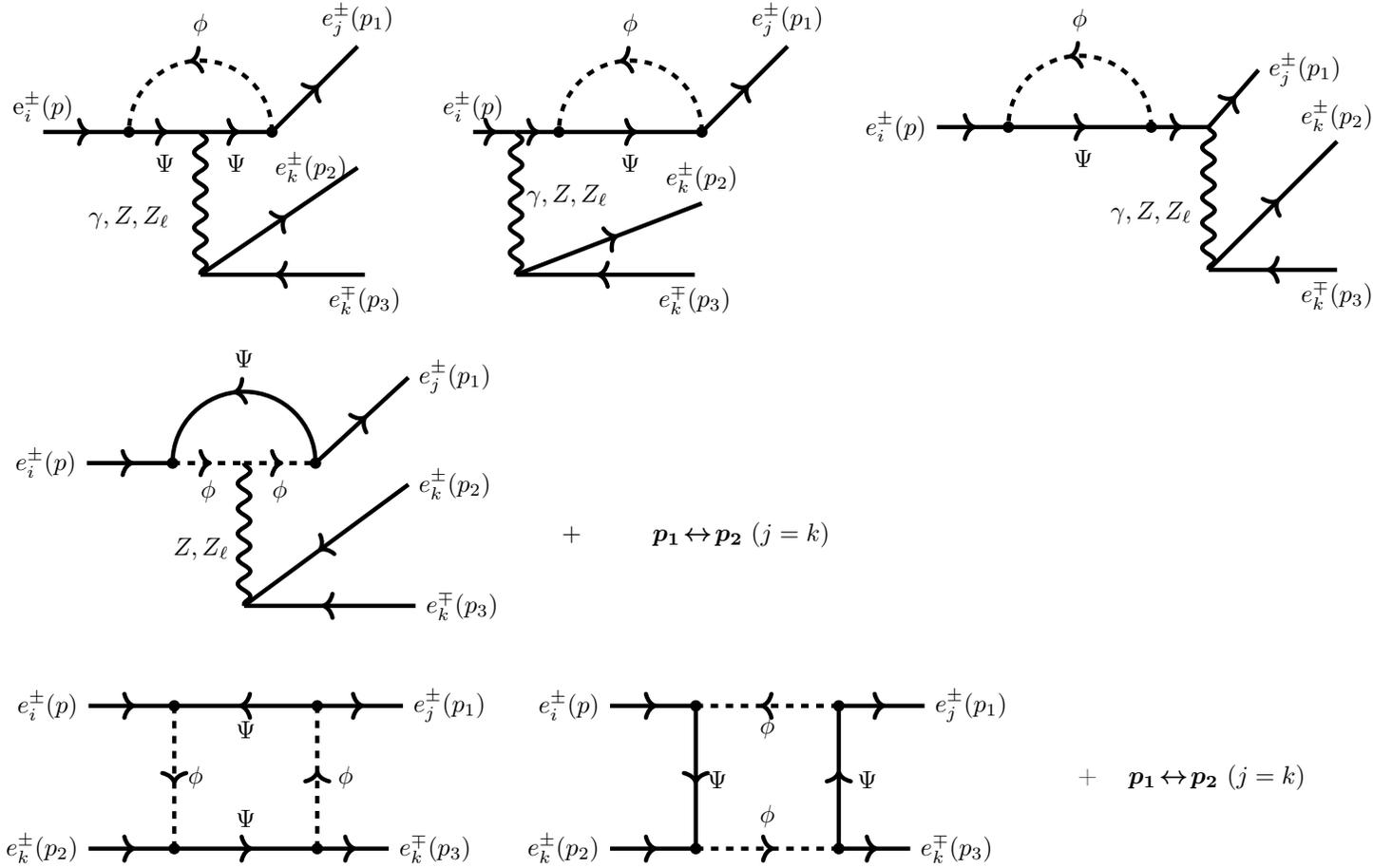
\begin{figure}[h]
\begin{eqnarray*}
\begin{gathered}
\begin{tikzpicture}[line width= 1.7 pt,node distance=1 cm and 1 cm]
\coordinate[label=above:$\e_i^{\pm}(p)$](a);
\coordinate[right = 1.2 cm of a](v1);
\coordinate[right = 2 cm of v1](v2);
\coordinate[right = 1 cm of v1](aux);
\coordinate[above = 1.2 cm of aux,label=$\phi$](phi);
\coordinate[right = 1.2 cm of v2](e);
\coordinate[above = 1.2 cm of e,label=above:$e_j^{\pm}(p_1)$](e1);
\coordinate[right = 1 cm of v1](psi);
\coordinate[below = 2 cm of psi](psi2);
\coordinate[below = 1.7 cm of e1,label=left:$e_k^{\pm}(p_2)$](q2);
\coordinate[right = 2.3 cm of psi2,label=below:$e_k^{\mp}(p_3)$](q3);
\coordinate[right = 0.5 cm of v1](psi3);
\coordinate[right = 1.5 cm of v1](psi4);
\coordinate[right = 1.8 cm of v1](g1);
\coordinate[below = 0.7 cm of psi3,label=$\Psi$](psi5);
\coordinate[below = 0.7 cm of psi4,label=$\Psi$](psi6);
\coordinate[below = 1.5 cm of v1,label={$\gamma,Z,Z_\ell$}](gamma);
\draw[fill=black](v1) circle (.05cm);
\draw[fill=black](v2) circle (.05cm);
\draw[fermion](a)--(v1);
\draw[fermion](v1)--(psi);
%\draw[fermionnoarrow](v1)--(v2);
\draw[fermion](psi)--(v2);
\draw[fermion](v2)--(e1);
\draw[vector](psi) -- (psi2);
\draw[fermion](q3)--(psi2);
\draw[fermion](psi2)--(q2);
\semiloop[scalar]{v1}{v2}{0};
\end{tikzpicture}
\end{gathered}  
\quad \! 
\begin{gathered}
\begin{tikzpicture}[line width= 1.7 pt,node distance=1 cm and 1 cm]
\coordinate[label=above:$e_i^{\pm}(p)$](a);
\coordinate[right = 1.2 cm of a](v1);
\coordinate[right = 2 cm of v1](v2);
\coordinate[right = 1 cm of v1](aux);
\coordinate[above = 1.2 cm of aux,label=$\phi$](phi);
\coordinate[right = 1.2 cm of v2](e);
\coordinate[above = 1.2 cm of e,label=above:$e_j^{\pm}(p_1)$](e1);
\coordinate[right = 0.6 cm of a](psi);
\coordinate[below = 2 cm of psi](psi2);
\coordinate[below = 1.0 cm of v2,label=above:$e_k^{\pm}(p_2)$](q2);
\coordinate[right = 2.5 cm of psi2,label=below:$e_k^{\mp}(p_3)$](q3);
\coordinate[right = 1 cm of v1](psi3);
\coordinate[right = 1.5 cm of v1](psi4);
\coordinate[right = 1.8 cm of v1](g1);
\coordinate[below = 0.7 cm of psi3,label=$\Psi$](psi5);
%\coordinate[below = 0.7 cm of psi4,label=$\Psi$](psi6);
\coordinate[below = 1.2 cm of v1,label={$ \ \ \gamma,Z,Z_\ell$}](gamma);
\draw[fill=black](v1) circle (.05cm);
\draw[fill=black](v2) circle (.05cm);
\draw[fermion](a)--(psi);
\draw[fermion](psi)--(v1);
\draw[fermion](v1)--(v2);
%\draw[fermionnoarrow](v1)--(v2);
%\draw[fermion](psi)--(v2);
\draw[fermion](v2)--(e1);
\draw[vector](psi) -- (psi2);
\draw[fermion](q3)--(psi2);
\draw[fermion](psi2)--(q2);
\semiloop[scalar]{v1}{v2}{0};
\end{tikzpicture}
\end{gathered}
\quad \!
\begin{gathered}
\begin{tikzpicture}[line width= 1.7 pt,node distance=1 cm and 1 cm]
\coordinate[label=left:$e_i^{\pm}(p)$](a);
\coordinate[right = 1 cm of a](v1);
\coordinate[right = 2 cm of v1](v2);
\coordinate[right = 1 cm of v1](aux);
\coordinate[above = 1.2 cm of aux,label=$\phi$](phi);
\coordinate[right = 1.5 cm of v2](e);
\coordinate[above = 0.8 cm of e,label=right:$e_j^{\pm}(p_1)$](e1);
\coordinate[right = 2.8 cm of v1](psi);
\coordinate[below = 2 cm of psi](psi2);
\coordinate[right = 1.8 cm of psi2](q4);
\coordinate[above = 1.8 cm of q4,label=above:$e_k^{\pm}(p_2)$](q2);
\coordinate[right = 1.8 cm of psi2,label=below:$e_k^{\mp}(p_3)$](q3);
\coordinate[right = 1.05 cm of v1](psi3);
\coordinate[right = 1.5 cm of v1](psi4);
\coordinate[right = 1.8 cm of v1](g1);
\coordinate[below = 0.7 cm of psi3,label=$\Psi$](psi5);
%\coordinate[below = 0.7 cm of psi4,label=$\Psi$](psi6);
\coordinate[below = 1.5 cm of v2,label={$\gamma,Z,Z_\ell$}](gamma);
\draw[fill=black](v1) circle (.05cm);
\draw[fill=black](v2) circle (.05cm);
\draw[fermion](a)--(v1);
\draw[fermion](v1)--(v2);
%\draw[fermionnoarrow](v1)--(v2);
\draw[fermion](v2)--(psi);
\draw[fermion](psi)--(e1);
\draw[vector](psi) -- (psi2);
\draw[fermion](q3)--(psi2);
\draw[fermion](psi2)--(q2);
\semiloop[scalar]{v1}{v2}{0};
\end{tikzpicture}
\end{gathered} \\
\quad \!
\begin{gathered}
\text{\hspace*{-2.7cm}\makebox[1.13\textwidth][l]{%
  \begin{tikzpicture}[line width=1.7pt,node distance=1 cm and 1 cm, baseline=(current bounding box.center)]
   \coordinate[label=left:$e_i^{\pm}(p)$](a);
\coordinate[right = 1.2 cm of a](v1);
\coordinate[right = 2 cm of v1](v2);
\coordinate[right = 1 cm of v1](aux);
\coordinate[above = 1.2 cm of aux,label=$\Psi$](phi);
%\coordinate[right = 1.2 cm of v2,label=right:$e1$](e);
\coordinate[above = 1.2 cm of e,label=right:$e_j^{\pm}(p_1)$](e1);
\coordinate[right = 1 cm of v1](psi);
\coordinate[below = 2 cm of psi](psi2);
\coordinate[below = 1.5 cm of e1,label=right:$e_k^{\pm}(p_2)$](q2);
\coordinate[right = 2.4 cm of psi2,label=right:$e_k^{\mp}(p_3)$](q3);
\coordinate[right = 0.5 cm of v1](psi3);
\coordinate[right = 1.5 cm of v1](psi4);
\coordinate[right = 0.4 cm of v1](g1);
\coordinate[below = 0.7 cm of psi3,label=$\phi$](psi5);
\coordinate[below = 0.7 cm of psi4,label=$\phi$](psi6);
\coordinate[below = 1.5 cm of g1,label={$Z,Z_\ell$}](gamma);
\draw[fill=black](v1) circle (.05cm);
\draw[fill=black](v2) circle (.05cm);
\draw[fermion](a)--(v1);
\draw[scalar](v1)--(psi);
%\draw[fermionnoarrow](v1)--(v2);
\draw[scalar](psi)--(v2);
\draw[fermion](v2)--(e1);
\draw[vector](psi) -- (psi2);
\draw[fermion](q2)--(psi2);
\draw[fermion](q3)--(psi2);
\semiloop[fermion]{v1}{v2}{0};
\node[anchor=west] at ([xshift=2cm ,yshift = -1 cm]e) {  $ + \hspace{1 cm} \bm{p_1\!\leftrightarrow\!p_2} \ (j=k)$};
  \end{tikzpicture}
}}
%\hspace{2cm}\raisebox{0.3em}{$p_1 \to p_2$}
\end{gathered}
\end{eqnarray*}\\
%\text{\vspace{5cm}}
%%%%%%%%%%%%%%%%%%%%%%%%%%%%%%
\begin{eqnarray*}
\hspace*{-0.1 cm}
\begin{gathered}
\begin{tikzpicture}[line width= 1.7 pt,node distance=1 cm and 1 cm]
\coordinate[label=left:$e_i^{\pm}(p)$](a);
\coordinate[right = 1.2 cm of a](v1);
\coordinate[right = 2 cm of v1](v2);
\coordinate[right = 1 cm of v1](aux);
%\coordinate[above = 1.2 cm of aux,label=$\phi$](phi);
\coordinate[right = 1.2 cm of v2,label=right:$e_j^{\pm}(p_1)$](e);
\coordinate[right = 1 cm of v1](psi);
\coordinate[below = 2 cm of psi](psi2);
\coordinate[left = 2.2 cm of psi2,label=left:$e_k^{\pm}(p_2)$](q2);
\coordinate[right = 2.0 cm of psi2,label=right:$e_k^{\mp}(p_3)$](q3);
\coordinate[below = 2 cm of v1](v3);
\coordinate[below = 2 cm of v2](v4);
\coordinate[right = 0.5 cm of v1](psi3);
\coordinate[right = 1.5 cm of v1](psi4);
\coordinate[right = 1.8 cm of v1](g1);
%\coordinate[below = 0.7 cm of psi3,label=$\Psi$](psi5);
%\coordinate[below = 0.7 cm of psi4,label=$\Psi$](psi6);
%\coordinate[below = 1.5 cm of g1,label={$\gamma,Z,Z_\ell$}](gamma);
\coordinate[right = 0.3 cm of v1](p1);
\coordinate[right = 0.4 cm of v2](p2);
\coordinate[right = 1 cm of v3](p3);
\coordinate[right = 0.4 cm of v3](p4);
\coordinate[below = 1.3 cm of p1,label=$\phi$](psi5);
\coordinate[below = 1.3 cm of p2,label=$\phi$](psi6);
\coordinate[above = 0.15 cm of p3,label=$\Psi$](psi7);
\coordinate[above = 1.4 cm of p3,label=$\Psi$](psi8);
\draw[fill=black](v1) circle (.05cm);
\draw[fill=black](v2) circle (.05cm);
\draw[fill=black](v3) circle (.05cm);
\draw[fill=black](v4) circle (.05cm);
\draw[fermion](a)--(v1);
\draw[fermion](v2)--(v1);
%\draw[fermionnoarrow](v1)--(v2);
%\draw[fermion](psi)--(v2);
\draw[fermion](v2)--(e);
%\draw[vector](psi) -- (psi2);
\draw[fermion](q2)--(v3);
\draw[fermion](v3)--(v4);
\draw[fermion](v4)--(q3);
\draw[scalar](v1)--(v3);
\draw[scalar](v4)--(v2);
%\semiloop[scalar]{v1}{v2}{0};
\end{tikzpicture}
\end{gathered} 
\quad \!
\begin{gathered}
\begin{tikzpicture}[line width= 1.7 pt,node distance=1 cm and 1 cm]
\coordinate[label=left:$e_i^{\pm}(p)$](a);
\coordinate[right = 1.2 cm of a](v1);
\coordinate[right = 2 cm of v1](v2);
\coordinate[right = 1 cm of v1](aux);
%\coordinate[above = 1.2 cm of aux,label=$\phi$](phi);
\coordinate[right = 1.2 cm of v2,label=right:$e_j^{\pm}(p_1)$](e);
\coordinate[right = 1 cm of v1](psi);
\coordinate[below = 2 cm of psi](psi2);
\coordinate[left = 2.2 cm of psi2,label=left:$e_k^{\pm}(p_2)$](q2);
\coordinate[right = 2.0 cm of psi2,label=right:$e_k^{\mp}(p_3)$](q3);
\coordinate[below = 2 cm of v1](v3);
\coordinate[below = 2 cm of v2](v4);
\coordinate[right = 0.5 cm of v1](psi3);
\coordinate[right = 1.5 cm of v1](psi4);
\coordinate[right = 1.8 cm of v1](g1);
\coordinate[right = 0.3 cm of v1](p1);
\coordinate[right = 0.4 cm of v2](p2);
\coordinate[right = 1 cm of v3](p3);
\coordinate[right = 0.4 cm of v4](p4);
\coordinate[below = 1.3 cm of p1,label=$\Psi$](psi5);
\coordinate[below = 1.3 cm of p2,label=$\Psi$](psi6);
\coordinate[above = 0.15 cm of p3,label=$\phi$](psi7);
\coordinate[above = 1.4 cm of p3,label=$\phi$](psi8);
%\coordinate[below = 0.7 cm of psi4,label=$\Psi$](psi6);
%\coordinate[below = 1.5 cm of g1,label={$\gamma,Z,Z_\ell$}](gamma);
\draw[fill=black](v1) circle (.05cm);
\draw[fill=black](v2) circle (.05cm);
\draw[fill=black](v3) circle (.05cm);
\draw[fill=black](v4) circle (.05cm);
\draw[fermion](a)--(v1);
\draw[scalar](v2)--(v1);
%\draw[fermionnoarrow](v1)--(v2);
%\draw[fermion](psi)--(v2);
\draw[fermion](v2)--(e);
%\draw[vector](psi) -- (psi2);
\draw[fermion](q2)--(v3);
\draw[scalar](v3)--(v4);
\draw[fermion](v4)--(q3);
\draw[fermion](v1)--(v3);
\draw[fermion](v4)--(v2);
%\semiloop[scalar]{v1}{v2}{0};
\node[anchor=west] at ([xshift=2 cm ,yshift = -1 cm]e) {   + \hspace{0.2 cm} $\bm{p_1\!\leftrightarrow\!p_2} \ (j=k)$};
\end{tikzpicture}
\end{gathered} 
\end{eqnarray*}
\caption{Feynman diagrams for $e_i^{\pm} \to e_j^{\pm} e_k^{\pm} e_{k}^{\mp}$. }
    \label{muto3e}
\end{figure}
%%%%%%%%%%%%%%%%%%%%%%%%%%%%%%%%%%%%%%%%%%%%%%%%%%%%%%%%%%%
The amplitude for the $e_i \to 3 e_j $ is given by 
\begin{eqnarray}
    i\mathcal{M}= \lambda_{e_j} \lambda_{e_i} \bar{u}_j(p_1)\left[q^2 \gamma^\nu A_1^R P_R + i m_\mu \sigma^{\nu \rho}q_\rho A_L P_L \right]u_i(p) \frac{e^2}{q^2} \bar{u}_j(p_2)\gamma_\nu u_j(p_3) - (p1 \leftrightarrow p2).
\end{eqnarray}
Here we take into account the most important contribution mediated by the photon.
The decay width for this process can be written as~\cite{Hisano:1995cp}
\begin{eqnarray}
    \Gamma(e_i\to 3e_j)= \frac{\alpha^2 \lambda_{e_i}^2 \lambda_{e_j}^2}{32 \pi}m_{e_i}^5 \Bigg[\left|A_1^R\right|^2 -2 \left(A_1^{R*}A_L+  A_1^{R} A_{L}^{*}\right)+ \left|A_L\right|^2 \left(\frac{16}{3} ln\left(\frac{m_{e_i}}{2m_{e_j}}\right)-\frac{14}{9}\right) \Bigg],
\end{eqnarray}
with $A_1^R = F_R^{(\gamma)}/(\lambda_{e_i} \lambda_{e_j} e)$. 
%%%%%%%%%%%%%%%%%%%%%%%%%%%%%%

\end{widetext}
\newpage
\bibliography{ref}

\end{document}